\documentclass[aps,prl,twocolumn,longbibliography,amssymb,10pt]{revtex4-1}
\usepackage{graphicx}
\usepackage{dcolumn}
\usepackage{bm}
\usepackage{natbib}
\usepackage[utf8x]{inputenc}
\usepackage{hyperref}
\usepackage[mathlines]{lineno}
\usepackage{color}
\usepackage{amsthm}
\usepackage{amsmath}
\usepackage{amssymb}
\usepackage{mathrsfs}
\usepackage{array}
\usepackage{graphicx}
\usepackage[normalem]{ulem}
\usepackage{verbatim,bm}
\usepackage[normalem]{ulem}
\usepackage{dsfont}

\usepackage{hyperref}

\newcommand*{\beginsupplement}{%
        \setcounter{table}{0}
        \renewcommand{\thetable}{S\arabic{table}}%
        \setcounter{figure}{0}
        \renewcommand{\thefigure}{S\arabic{figure}}%
        \setcounter{equation}{0}
        \renewcommand{\theequation}{S\arabic{equation}}%
     }

\makeatother

\begin{document}
\preprint{APS/123-QED}

\title{Thermodynamic consistency of quantum master equations}

\author{Ariane Soret$^{\dagger,*}$, Vasco Cavina$^\dagger$, Massimiliano Esposito$^\dagger$}

\affiliation{$^\dagger$Complex Systems and Statistical Mechanics, Department of Physics and Materials Science, University of Luxembourg, L-1511 Luxembourg, Luxembourg}

\email{ariane.soret@gmail.com}

\date{\today}

\begin{abstract}
Starting from a microscopic system-baths description, we derive the general conditions for a time-local quantum master equation (QME) to satisfy the first and second law of thermodynamics at the fluctuating level. Using counting statistics, we show that the fluctuating second law can be rephrased as a Generalized Quantum Detailed Balance condition (GQDB), i.e., a symmetry of the time-local generators which ensures the validity of the fluctuation theorem.
When requiring in addition a strict system-bath energy conservation, the GQDB reduces to the usual notion of detailed balance which ensures QMEs with Gibbsian steady states. However, if energy conservation is only required on average, QMEs with non Gibbsian steady states can still maintain a certain level of thermodynamic consistency.
Applying our theory to commonly used QMEs, we show that the Redfield equation breaks the GQDB, and that some recently derived approximation schemes based on the Redfield equation (which hold beyond the secular approximation and allow to derive a QME of Lindblad form) satisfy the GQDB and the average first law. 
We find that performing the secular approximation is the only way to ensure the first and second law at the fluctuating level.
\end{abstract}

\maketitle


Quantum master equations (QMEs) are widely used to describe the dynamics of open quantum systems.
The issue of their thermodynamic consistency has been raised early on, and has in general been tackled by assuming the validity of detailed balance conditions, which impose strict constraints on the generator of the QME and on the associated steady state density matrix \cite{alicki1976detailed,gorini1977detailed,spohn1978irreversible,fagnola2007generators,talkner1986failure, agarwal1973open,majewski1984detailed}. 
The rapid development of quantum and stochastic thermodynamics \cite{seifert2012stochastic, strasberg2021} in recent years has created a renewed interest in this topic and many efforts have been devoted to identifying conditions under which QMEs satisfy quantum analogs of classical fluctuation theorems (FT) \cite{esposito2007ftelectron,esposito2009nonequilibrium,campisi2011colloquium,cuetara2016quantumthermo,chetrite2012quantum,manzano2018quantum}.
For systems in contact with one or more environments initialized at equilibrium, FTs emerge as exact symmetries governing the fluctuations of thermodynamic quantities (such as work and heat currents) at the level of the unitary evolution \cite{kurchan2000quantum,andrieux2009fluctuation}.
A thermodynamically consistent QME, obtained by tracing out the baths, should therefore preserve such symmetries. In this letter, we show that FTs at the unitary level translate in a condition that the QMEs \emph{must} satisfy in order to be thermodynamically consistent -- the Generalized Quantum Detailed Balance condition (GQDB). Combined with a strict energy conservation at the fluctuating level, the GQDB reduces to the usual detailed balance. 
However, this strict condition is not necessary to ensure the first and second law at the average level. The criteria derived here are used to discuss the consistency of usual microscopic derivations w.r.t. the GQDB.
We show that the latter is violated by the Redfield QME, and discuss whether the approximation schemes used to restore the Lindblad form of the Redfield ME \cite{wacker2018,ptaszynski2019thermodynamics,farina2019,mccauley2020accurate,nathan2020universal, potts2021} allow to recover the GQDB. Imposing also a strict energy conservation, that is, requiring consistency at the fluctuating level, requires to perform the secular approximation.

\emph{Thermodynamics at the system-baths unitary level:}
Let us consider a system $S$ coupled to $N$ baths, labelled by $\alpha=1,2,..,N$.
The total Hamiltonian reads
\begin{align} \hat{H}(t)=\hat{H}_S(t)+\sum_{\alpha=1}^N \hat{H}_\alpha + \sum_{\alpha=1}^N \hat{V}_\alpha(t), \label{eq:ham} \end{align} 
%
where $\hat{H}_S(t)$ is the system Hamiltonian, $\hat{H}_\alpha$ the free Hamiltonian of the $\alpha$-th bath and $\hat{V}(t)=\sum_{\alpha=1}^N\hat{V}_\alpha(t)$ the coupling Hamiltonian between the baths and the system. The moment generating function (MGF) of the probability distribution of energy changes resulting from a double projective measurement of the operators $\{\hat{H}_S, \hat{H}_{\alpha}\}$ at times $0$ and $t$ is given by \cite{esposito2009nonequilibrium}
\begin{equation}
\begin{array}{ll}
G(t,\boldsymbol{\lambda})= Tr[\hat{\rho}_{\boldsymbol{\lambda}}(t)];  \quad \hat{\rho}_{\boldsymbol{\lambda}}(t)
=\hat{U}_{\boldsymbol{\lambda}}(t,0)\bar{\hat{\rho}}_0 \hat{U}^\dagger_{\boldsymbol{\lambda}}(t,0),
\end{array}
\label{eq:mgf}
\end{equation}
where
$\hat{U}_{\boldsymbol{\lambda}}(t,0)$ is the tilted evolution operator,
\begin{equation} 
\begin{array}{ll}
\hat{U}_{\boldsymbol{\lambda}}(t,0)=
e^{\boldsymbol{\lambda}\cdot\boldsymbol{\hat{H}}/2} \hat{U}(t,0) e^{-\boldsymbol{\lambda}\cdot\boldsymbol{\hat{H}}/2} ,&
\end{array}
\label{eq:Utilt} \end{equation}
with $\boldsymbol{\hat{H}}=(\hat{H}_S,\hat{H}_1,...,\hat{H}_N)$, counting field $\boldsymbol{\lambda}=(\lambda_S,\lambda_1,...,\lambda_N)$ and $\bar{\hat{\rho}}_0$ denoting the diagonal part of the initial density matrix in the basis diagonalizing \eqref{eq:ham} in absence of coupling. 
Fluctuations of various quantities can be obtained from the MGF by a suitable choice of the counting field $\boldsymbol{\lambda}$. For the system bare energy changes, $\Delta E_S$, we choose $\lambda_S = \lambda$ and $\lambda_\alpha=0$ for all $\alpha$. 
For the energy changes leaving bath $\eta$, i.e., the heat $Q_\eta$, we choose $\lambda_S=\lambda_\alpha=0$ for all $\alpha\neq\eta$ and $\lambda_\eta=-\lambda$. 
Finally, for the ``work'', $W = \Delta E_S-\sum_{\eta} Q_\eta$, we choose $\lambda_S=\lambda_\alpha=\lambda$. 
We note however that the term work is only justified when additionally requiring that the coupling is switched on/off after/before the initial/final measurement:  $\hat{V}_{\alpha}(0)=\hat{V}_{\alpha}(t)=0$. Otherwise, this coupling contribution should be included into the system energy \cite{esposito2010entropy}. 
Detailed FTs are exact symmetries of the MGF, connecting the fluctuation in a forward process to its time reversed process. The MGF for the reversed dynamics is given by \cite{esposito2009nonequilibrium}
\begin{equation}
\begin{array}{ll}
G^{R}(t,\boldsymbol{\lambda})= Tr[\hat{\rho}^R_{\boldsymbol{\lambda}}(t)]
=Tr\left[\hat{U}_{\boldsymbol{-\lambda}}^{\dagger}(t,0)\bar{\hat{\rho}}_0^R \hat{U}_{\boldsymbol{-\lambda}}(t,0)\right].
\end{array}
\label{eq:mgfrev}
\end{equation}
If the initial density matrices of the forward and backward processes are respectively $\hat{\rho}_0= \frac{e^{- \beta_S \hat{H}_S(0)}}{Z_S(0)} \bigotimes_\alpha \frac{e^{- \beta_\alpha \hat{H}_\alpha}}{Z_\alpha}$ and $\hat{\rho}_0^R=\frac{e^{-\beta_S \hat{H}_S(t)}}{Z_S(t)}\bigotimes_\alpha \frac{e^{- \beta_\alpha \hat{H}_\alpha}}{Z_\alpha}$, with $Z_S(t)=Tr_S[e^{- \beta_S \hat{H}_S(t)}]$ and $Z_\alpha=Tr_\alpha[e^{- \beta_\alpha \hat{H}_\alpha}]$, 
%
%
the following detailed FT can be derived (see \cite{suppmat}, section 1)
\begin{equation}
    G^{R}(t,\boldsymbol{-\lambda-\beta})=G(t,\boldsymbol{\lambda})\frac{Z_S(0)}{Z_S(t)}=G(t,\boldsymbol{\lambda})e^{\beta_S\Delta F_{eq}},
    \label{eq:sym-uni-W}
\end{equation}
with inverse temperatures $\boldsymbol{\beta}=(\beta_S,\beta_1,...,\beta_N)$ ($k_B=1$) and $\Delta F_{eq}=F_{eq}(t)-F_{eq}(0)$ where $F_{eq}=-\frac{1}{\beta_S}\log Z_S(t) $ is the equilibrium free energy of the system. 
A related but in general different FT can be derived for the entropy production $\Sigma$, corresponding to changes in the quantity $\hat{S}(t)+\sum_\alpha \beta_\alpha \hat{H}_\alpha$, where $\hat{S}(t) = -\log\hat{\rho}_S(t) $ is the system entropy. This time, only the initial density matrices of the baths need to be in a Gibbs state and we obtain (see \cite{suppmat}, section 1)
\begin{equation}
    G^{R}_{\Sigma}(t,-\lambda_{\Sigma}-1)=G_{\Sigma}(t,\lambda_{\Sigma}).
    \label{eq:sym-uni-entropy}
\end{equation}
As such this relation is rather formal, but setting $\lambda_\Sigma=-1$, we obtain the integral fluctuation theorem, $G_{\Sigma}(t,-1) = \langle e^{-\Sigma} \rangle = 1$ \footnote{The brackets denote an ensemble averaging; see also \cite{suppmat}, section 1},
which, using the convexity of the exponential function, implies the second law $ \langle \Sigma \rangle \geq 0$ \cite{seifert2005entropy,jarzynski1997nonequilibrium} (see \cite{suppmat}, section 1). Notice that, if the system is also prepared in a Gibbs state, \eqref{eq:sym-uni-entropy} becomes a special case of \eqref{eq:mgfrev}. Finally, note that when the Hamiltonians \eqref{eq:ham} are time-independent, the requirement that fluctuations in $W$ vanish -- i.e. that system-baths energy is conserved at the fluctuating level --  implies the invariance 
\begin{equation}
    \hat{\rho}_{\boldsymbol{\lambda} +\chi \boldsymbol{1}}(t) = \hat{\rho}_{\boldsymbol{\lambda}}(t), \label{eq:invtrasl}
\end{equation}
for all times, where $\boldsymbol{1} = (1,1...,1)$ and $\chi\in\mathbb{R}$.
One easily verifies that this strict energy conservation holds true when $[\hat{V}, \hat{H}_S + \sum_{\alpha} \hat{H}_{\alpha}]=0$. 

%
\emph{Thermodynamics at the QME level:}
We now proceed to show how the conditions \eqref{eq:sym-uni-W}-\eqref{eq:invtrasl} translate at the level of effective QMEs after tracing out the bath degrees of freedom -- a procedure we will also refer to as coarse graining.
Decomposing the initial state as $\hat{\rho}_0 = \hat{\rho}_S(0) \otimes \sum_{\nu} \eta_{\nu} | \nu \rangle \langle \nu|$, where $| \nu \rangle$ are eigenvectors of the baths, the tilted density matrix in \eqref{eq:mgf} becomes
\begin{equation}  \hat{\rho}_S^{\boldsymbol{\lambda}}(t) =  \sum_{\mu,\nu} \hat{W}_{\mu,\nu}^{\boldsymbol{\lambda}} (t,0) \hat{\rho}_S(0)
\hat{W}_{\mu,\nu}^{\boldsymbol{\lambda}\, \dagger}(t,0) \equiv \hat{M}_{\boldsymbol{\lambda}}(t,0)\hat{\rho}_S(0), \label{eq:genopen} \end{equation}
where $\hat{W}_{\mu,\nu}^{\boldsymbol{\lambda}} = \sqrt{\eta_{\nu}} \langle \mu | \hat{U}_{\boldsymbol{\lambda}}(t,0)  |\nu \rangle$ are a set of Kraus operators acting on the system Hilbert space \cite{breuer2002theory}. 
The Markov approximation is enforced by assuming the semigroup hypothesis: $\hat{M}_{\boldsymbol{\lambda}}(t,0)=\hat{M}_{\boldsymbol{\lambda}}(t,s)\hat{M}_{\boldsymbol{\lambda}}(s,0)$, leading to a time local equation of the form
\begin{equation} d_t \hat{\rho}_{S}^{\boldsymbol{\lambda}} (t) = \lim_{\delta\to \delta_0}\frac{1}{\delta}(\hat{M}_{\boldsymbol{\lambda}}(t+\delta,t)-\mathds{I})\hat{\rho}_{S}^{\boldsymbol{\lambda}}(t)\equiv
\mathcal{L}_{\boldsymbol{\lambda}}(t) \hat{\rho}_{S}^{\boldsymbol{\lambda}}(t), \label{eq:teorlind} \end{equation}
with $\hat{\rho}_{S}^{\boldsymbol{\lambda}}(0) = \hat{\rho}_S(0)$, and where the coarse graining time $\delta_0$ needs to be simultaneously larger than the typical correlation time of the bath and smaller than the relaxation time of the system. Throughout this paper, we assume that the limit in \eqref{eq:teorlind} exists and does not explicitly depend on $\delta_0$.
As shown in \cite{suppmat}, section 2, when counting the energy flow to the thermal baths $\boldsymbol{\lambda}_B= 
(\lambda_1,...\lambda_N)$, the generator assumes the general form \footnote{To alleviate the notation, we dropped the $t$ dependence of $\hat{\rho}_S^{0,\boldsymbol{\lambda_B}}(t)$.}
\begin{equation}  \mathcal{L}_{0,\boldsymbol{\lambda}_B}(t) \hat{\rho}_S^{0,\boldsymbol{\lambda}_B} = - i [\hat{H}_{0,\boldsymbol{\lambda_B}}'(t), \hat{\rho}_S^{0,\boldsymbol{\lambda}_B}] +  \mathcal{D}_{0,\boldsymbol{\lambda}_B}(t) \hat{\rho}_S^{0,\boldsymbol{\lambda}_B},\label{eq:dynlind}
\end{equation}
where $\hat{H}_{0,\boldsymbol{\lambda_B}}'(t) $ is identified as the sum of the system Hamiltonian $\hat{H}_S$ and of a Lamb shift contribution  $\hat{H}^{0,\boldsymbol{\lambda_B}}_{LS}$, 
and $\mathcal{D}_{0,\boldsymbol{\lambda}_B}(t)$ accounts for the dissipation. 
The dissipator itself can be decomposed in an anticommutator 
and a jump term $\mathcal{J}$, 
\begin{equation}
\mathcal{D}_{0,\boldsymbol{\lambda}_B}(t)\hat{\rho}_S^{0,\boldsymbol{\lambda}_B} = \{\hat{G}_{0,\boldsymbol{\lambda}_B},\hat{\rho}_S^{0,\boldsymbol{\lambda}_B}\} + \mathcal{J}_{0,\boldsymbol{\lambda}_B}(t) \hat{\rho}_S^{0,\boldsymbol{\lambda}_B}.
\label{eq:com-jump}
\end{equation}
The analog of \eqref{eq:genopen} for the time reversed MGF \eqref{eq:mgfrev} is given in terms of the Kraus operators $\hat{W}_{\mu,\nu}^{R\boldsymbol{\lambda}} = \sqrt{\eta_{\nu}} \langle \mu | \hat{U}^\dagger_{\boldsymbol{-\lambda}}(t,0)  |\nu \rangle$. As shown in \cite{suppmat}, section 2, these operators inherit a symmetry similar to \eqref{eq:sym-uni-W},
\begin{equation}
    \hat{W}_{\mu,\nu}^{R\,\boldsymbol{\lambda}}= e^{\lambda_S \hat{H}_S(0) }(\hat{W}_{\nu,\mu}^{(\lambda_S, -\boldsymbol{\lambda}_B -\boldsymbol{\beta}_B)} )^{\dagger} e^{-\lambda_S \hat{H}_S(t) } \, .
    \label{eq:symmW}
\end{equation}
Combined with the semigroup hypothesis, this relation leads to the following GQDB condition
\begin{equation}
    \mathcal{L}^{R}_{0,\boldsymbol{\lambda}_B}[...] = \mathcal{L}^{\dagger}_{0,-\boldsymbol{\lambda}_B - \boldsymbol{\beta}_B}[...],
    \label{eq:detailed}
\end{equation}
where $\boldsymbol{\beta}_B=(\beta_1,...,\beta_N)$, and where we introduced the adjoint $\mathcal{O}^{\dagger}$ of a superoperator $\mathcal{O}$ as the one fulfilling $Tr[(\mathcal{O}(X))^\dagger Y]=Tr[X^\dagger \mathcal{O}^\dagger(Y)] $ for all operators $X,Y$. 
Crucially, \eqref{eq:detailed} is a sufficient condition for the QME to satisfy the FTs \eqref{eq:sym-uni-W} and \eqref{eq:sym-uni-entropy} (see \cite{suppmat}, section 3).
%
%

We now turn to energy conservation and assume for simplicity that $\hat{H}_S$ is time-independent, although the upcoming discussion could be extended to driven systems, under certain conditions discussed in the conclusion. From the property of the Kraus operators $\hat{W}_{\mu\nu}^{\boldsymbol{\lambda}}=e^{\lambda_S\hat{H}_S/2}\hat{W}_{\mu\nu}^{0,\boldsymbol{\lambda_B}}e^{-\lambda_S\hat{H}_S/2}$ , we obtain, after taking the limit in \eqref{eq:teorlind}, the general form of a tilted superoperator,
\begin{align}
    \mathcal{L}_{\lambda_S,\boldsymbol{\lambda}_B}[...] &= \label{eq:add-ls} \\ &\hspace{-0.5cm} e^{\frac{\lambda_S}{2} \hat{H}_S} \mathcal{L}_{0,\boldsymbol{\lambda}_B}[e^{-\frac{\lambda_S }{2} \hat{H}_S} ...  e^{-\frac{\lambda_S}{2} \hat{H}_S}] e^{\frac{\lambda_S}{2} \hat{H}_S } \, .
    \nonumber
\end{align}
The condition \eqref{eq:invtrasl} then translates in the following strict energy balance condition
\begin{equation} \mathcal{L}_{\boldsymbol{\lambda}}[...] =  
\mathcal{L}_{\boldsymbol{\lambda}+ \chi \boldsymbol{1}}[...]. \label{eq:loccons}\end{equation}
This relation unveils a crucial point: there is a strict connection between invariance properties and symmetries of the generating function \cite{rao2018conservation,rao2018detailed, polettini2016conservation}. Notice that \eqref{eq:loccons} can be understood as imposing \eqref{eq:invtrasl} only over time intervals $\delta_0$, but not at all times. It is therefore less restrictive than \eqref{eq:invtrasl}.

\emph{Detailed balance and GQDB:}
Combining our GQDB condition  \eqref{eq:detailed} with the energy balance \eqref{eq:loccons} allows us to connect with the notion of detailed balance commonly used in the literature \cite{breuer2002theory}. 
Indeed, starting from \eqref{eq:detailed} for a single bath, setting $\lambda_B=0$, we can manipulate the r.h.s. using \eqref{eq:loccons} to obtain
\begin{equation}
     \mathcal{L}^{R}_{0,0}[...] = \mathcal{L}^{\dagger}_{0, - \beta_1}[...] =
     \mathcal{L}^{\dagger}_{\beta_1,0}[...], \label{eq:det1}
\end{equation}
from which we recover the usual detailed balance
\begin{equation}\label{eq:det-lit}
\mathcal{L}^{R}_{0,0}[...] = e^{-\frac{\beta_1}{2} \hat{H}_S } \mathcal{L}_{0,0}^\dagger[e^{\frac{\beta_1 }{2} \hat{H}_S } ...  e^{\frac{\beta_1}{2} \hat{H}_S}] e^{-\frac{\beta_1}{2} \hat{H}_S }\, .
\end{equation}
Note that quantum detailed balance conditions are often expressed in reference to a fixed point $\hat{\rho}_{ss}$, which in the case of a single bath is Gibbsian. Here, instead, \eqref{eq:detailed} does not depend on the steady state. However, assuming \eqref{eq:loccons} leads to \eqref{eq:det-lit}, which implies that the Gibbs state is a fixed point. This is easily proven by combining $\mathcal{L}^\dagger[\mathds{I}]=0$ \footnote{This follows from the fact that $\mathcal{L}$ is trace preserving and from the definition of the adjoint superoperator.} with \eqref{eq:det-lit}  to obtain $\mathcal{L}^R[e^{-\beta_1 \hat{H}_S}] =0$, and hence the Gibbs state is a steady state of $\mathcal{L}$ by virtue of the relation $\mathcal{L}^R[...]=\Theta\mathcal{L}[\Theta ...\Theta]\Theta$, where $\Theta$ is the time reversal operator satisfying $\Theta^2=\mathds{I}$ \cite{esposito2009nonequilibrium} (see \cite{suppmat}, section 4 for details).
The relation to other notions of detailed balance \cite{fagnola2007generators,fagnola2008detailed} is discussed further in \cite{suppmat}, section 5. 
The relation \eqref{eq:det1} can be trivially extended to multiple baths: since the coupling term in the Hamiltonian \eqref{eq:ham} is additive in the baths, the same holds true for the dissipators \eqref{eq:com-jump}, which will individually satisfy \eqref{eq:det1} with their respective temperatures.

\emph{Thermodynamics at the average level:}
The first (resp. second) law at the average level follows naturally from the energy conservation condition \eqref{eq:loccons} (resp. from the GQDB \eqref{eq:detailed}). Indeed, setting $\boldsymbol{\lambda=0}$ in \eqref{eq:loccons} implies that adding the same counting field $\chi$ on $\hat{H}_S$ and $\hat{H}_B$ leaves the superoperator invariant. Consequently, the work MGF, $G(t, \chi\boldsymbol{1})=Tr[\hat{\rho}_S^{\chi\boldsymbol{1}}]$, is $\chi$ independendent. Since, by linearity, $\partial_\chi G(t, \chi\boldsymbol{1})=\partial_\chi G(t, \lambda_S=\chi)+\sum_\alpha\partial_\chi G(t, \lambda_\alpha=\chi)$ -- where the two terms on the r.h.s. are respectively the MGF of the changes in the system energy $E_S(t)=Tr_S[ \hat{H}_S \rho_S(t)]$ 
and of the heat exchanges $Q_\alpha$ -- we obtain the first law
\begin{equation}
  \partial_\chi G(t, \chi\boldsymbol{1})|_{\chi=0}=  \Delta E_S -\sum\limits_{\alpha} Q_\alpha =0 \, .
  \label{eq:1stlaw}
\end{equation}
For the second law, since we already showed that the GQDB \eqref{eq:detailed} guarantees the FT \eqref{eq:sym-uni-entropy}, we can follow the same reasoning as for the unitary case and obtain, at the QME level, that $G_{\Sigma}(t,-1)=1$, hence
\begin{equation}
    \langle \Sigma\rangle=\Delta S-\sum\limits_\alpha\beta_\alpha Q_\alpha \geq 0,
    \label{eq:2ndlaw}
\end{equation}
where $S(t)=-Tr_S[\hat{\rho}_S(t) \ln \hat{\rho}_S(t)]$.
Using the fact that the semigroup (or Markov) approximation assumes that the generator is the same for every time interval, including the initial one calculated with the factorized initial condition, \eqref{eq:2ndlaw} holds true at the rate level. For the first and second law, we thus get
\begin{equation}
    d_t E_S = \sum\limits_\alpha \dot{Q}_\alpha \  \ \;, \  \ 
    \langle \dot{\Sigma}\rangle=d_t S-\sum\limits_\alpha\beta_\alpha \dot{Q}_\alpha \geq 0.
    \label{eq:2ndlawrate}
\end{equation}
For a single bath, $\dot{Q}=Tr_S [\hat{H}_S d_t\rho_S(t)]$ and thus $\langle \dot{\Sigma}\rangle=-d_t D(\rho_S(t)||e^{- \beta \hat{H}_S}/Z_S) \geq 0$. Here $D$ denotes the relative entropy which is a Lyapunov function of the dynamics since, as showed under \eqref{eq:det-lit}, the steady state is of a Gibbs form \cite{spohn1978irreversible,breuer2002theory}.


\emph{Thermodynamic consistency when deriving QMEs:}
In the light of the above, we now examine the consistency of microscopic derivations used in the literature. The Redfield ME, which
describes the open system dynamics under the Born-Markov approximation \cite{breuer2002theory}, violates 
\eqref{eq:detailed}. 
This can be seen by computing the associated jump term \eqref{eq:com-jump}, 
which is equal to
\begin{align}
&\mathcal{J}^{Red}_{0,\boldsymbol{\lambda_B}}(t)\hat{\rho}_S = \label{eq:diss-red-lambda}\\
&\sum\limits_{mn,m'n',\alpha}\big[(\Gamma^\alpha_+(\omega_{m'n'},\lambda_\alpha)^*+\Gamma^\alpha_+(\omega_{mn},\lambda_\alpha))\hat{A}_{mn}\hat{\rho}_S \hat{A}_{m'n'}^\dagger
\nonumber \\
&\hspace{1cm}+(\Gamma^\alpha_-(\omega_{m'n'},\lambda_\alpha)^*+\Gamma^\alpha_-(\omega_{mn},\lambda_\alpha))\hat{A}^\dagger_{mn}\hat{\rho}_S \hat{A}_{m'n'}\big], \nonumber
\end{align}
where $\hat{A}_{mn}=g_{mn}\hat{\sigma}_{mn}$, with $g_{mn}\in\mathbb{C}$ a complex amplitude and the operators $\{\hat{\sigma}_{mn}\}$ forming a basis of jump operators acting on the system Hilbert space. By definition, $\hat{\sigma}_{mn}=|E_n\rangle\langle E_m|$, where $\{|E_n\rangle\}$ are eigenvectors of $\hat{H}_S$ and $\omega_{mn}=E_m-E_n$ are the corresponding Bohr frequencies. $\Gamma^\alpha_\pm(\omega_{mn},\lambda_\alpha)$ are the one-sided Fourier transforms 
of the tilted baths correlation functions. They satisfy the relation $\Gamma^\alpha_\pm(\omega_{mn},-\lambda_{\alpha}-\beta_{\alpha})=\Gamma^\alpha_\mp(\omega_{mn},\lambda_{\alpha})^*$, from which we can immediately check that \eqref{eq:diss-red-lambda} does not satisfy \eqref{eq:detailed}  (see \cite{suppmat}, section 6 for details). 


%

%
\begin{figure}
    \centering
    \includegraphics[scale=0.6]{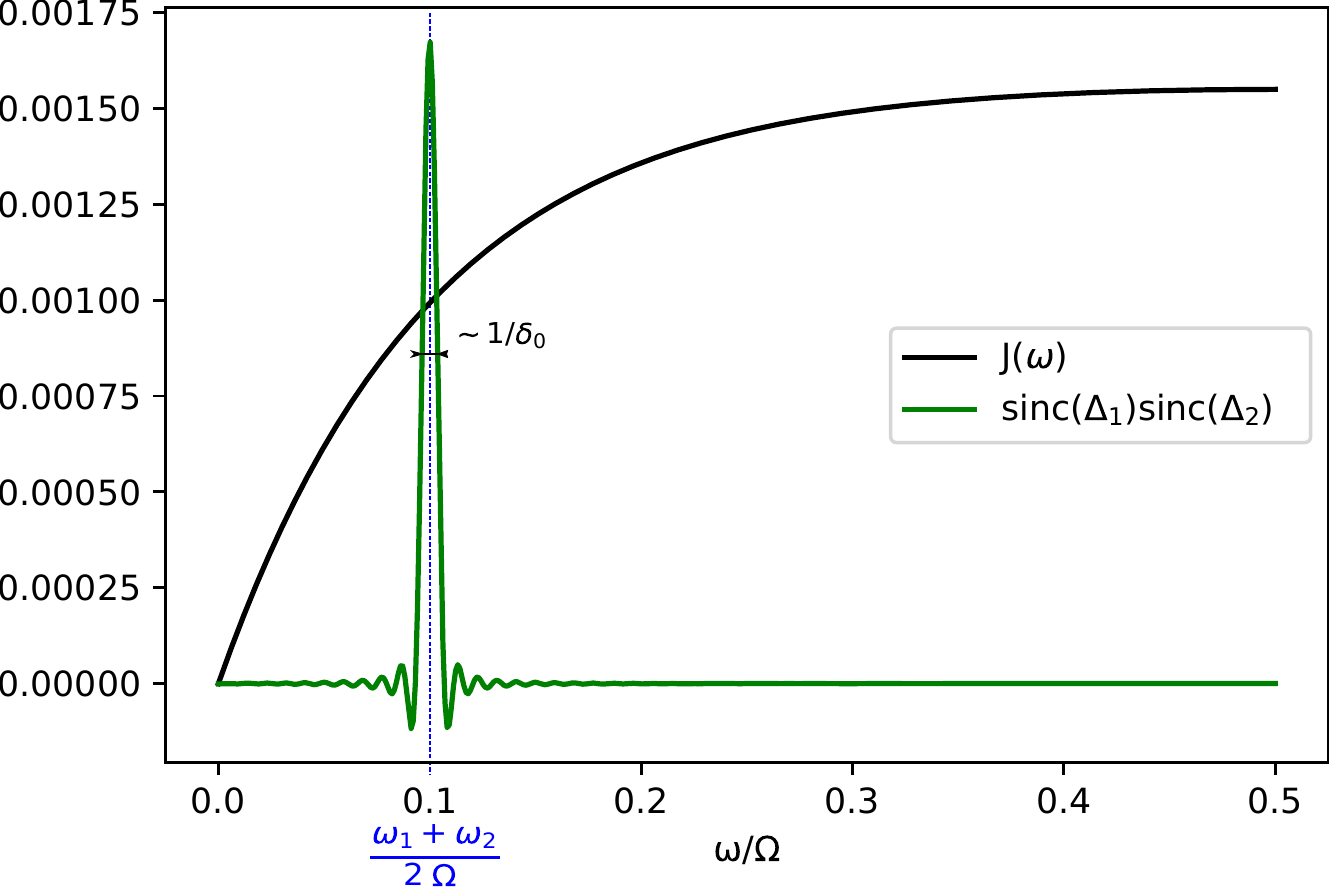}
    \caption{The product of the sinc functions of \eqref{eq:cond-1stlaw2} is represented in green, with $\Delta_{1,2}=(\omega_{1,2}-\omega)\delta_0/2$ and $\omega_2=\omega_1+1/\delta_0$. The black curve is the spectral density. See \cite{suppmat}, section 9 for details on the model and parameters. }
    \label{fig:sincs}
\end{figure}
To get a better understanding of the implications that the approximations made when deriving tilted QMEs of the form \eqref{eq:dynlind} have on the thermodynamics, we start directly from \eqref{eq:teorlind}.
%
In the basis $\{\hat{\sigma}_{mn}\}$ introduced earlier, the operator $\hat{M}_{\boldsymbol{\lambda}}$ takes the form $\sum d_{mn,m'n'}^{\boldsymbol{\lambda}}(t)\hat{\sigma}_{mn}[...]\hat{\sigma}_{m'n'}$, where $d_{mn,m'n'}^{\boldsymbol{\lambda}}(t)$ is
\begin{eqnarray}
&&\sum_{\mu,\nu} Tr_S[\hat{\sigma}_{mn}^\dagger\hat{W}^{\boldsymbol{\lambda}}_{\mu,\nu}(t+\delta,t)] Tr_S[\hat{\sigma}_{m'n'}^\dagger\hat{W}^{\boldsymbol{\lambda}}_{\mu,\nu}(t+\delta,t)]^*  \label{eq:coef-en-cons}\\
 &&=\eta_{\nu} e^{\frac{\lambda}{2}[2(\nu-\mu)- \omega_{nm}-\omega_{n'm'}]}\langle E_{n} | \hat{W}^{\boldsymbol{0}}_{\mu,\nu}  |E_{m}\rangle \langle E_{n'} | \hat{W}^{\boldsymbol{0}}_{\mu,\nu}|E_{m'}\rangle^*  \, , \nonumber
\end{eqnarray}
where $\hat{W}^{\boldsymbol{0}}_{\mu,\nu}$ is the Kraus operator for $\boldsymbol{\lambda}= \boldsymbol{0}$ (we dropped the $t$ dependence to alleviate the notation).
Strict energy conservation \eqref{eq:loccons} is achieved when the above term is $\lambda$ independent, which forces the allowed transitions to satisfy $2(\nu-\mu)= \omega_{nm}+\omega_{n'm'}$. This leads to $\nu-\mu = \omega_{nm} = \omega_{n'm'}$, since the two transition amplitudes in \eqref{eq:coef-en-cons} are independent. Hence, satisfying \eqref{eq:loccons} requires to perform the secular approximation. Since this procedure preserves the property \eqref{eq:symmW} of the Kraus operators, GQDB is preserved. Consequently, performing the secular approximation is the only way to guarantee both \eqref{eq:loccons} and \eqref{eq:detailed}.

To go further, we relax the energy conservation condition from the fluctuating level to the average level only.
Average energy conservation only requires the derivative in $\lambda$ of \eqref{eq:coef-en-cons} to vanish in 0, which in principle allows to keep more terms while preserving the symmetry \eqref{eq:symmW}. To illustrate this idea, consider the weak coupling limit, where we can pertubatively expand the propagator in $\hat{W}^{\boldsymbol{0}}_{\mu,\nu}$, and obtain the following condition for any $\omega_{nm},\omega_{n'm'}$ and for all $\alpha$ (see details in \cite{suppmat}, section 7):
\begin{align}
\int d\omega R^\alpha_\pm(\omega)\Delta_{nm} ^{n'm'}(\omega) \mbox{sc}\left(\frac{(\omega-\omega_{nm})\delta_0}{2}\right)&\mbox{sc}\left(\frac{(\omega-\omega_{n'm'})\delta_0}{2}\right) \nonumber\\
&=0,
\label{eq:cond-1stlaw2}
\end{align}
where $R^\alpha_\pm(\omega)=\Re(\Gamma^\alpha_\pm(\omega,0))$ are the Fourier transforms of the baths correlation functions, without counting fields, $\Delta_{mn}^{m'n'}(\omega) =\omega-(\omega_{mn}+\omega_{m'n'})/2$ and $\mbox{sc}$ is a short hand for $sinc$.  
This condition is obviously satisfied in the secular ($|\omega_{mn}-\omega_{m'n'}|\gg \delta_0^{-1}$) and degenerate ($\omega_{mn}=\omega_{m'n'}$) cases. 
But it also holds when the frequencies are nearly degenerate, $0<|\omega_{mn}-\omega_{m'n'}|< \delta_0^{-1}$. In this case, the product of $sinc$ functions is peaked around $(\omega_{mn}+\omega_{m'n'})/2$ (see Fig.\ref{fig:sincs}), hence, if $R^\alpha(\omega)$ is sufficiently smooth, the prefactor $\Delta_{mn}^{m'n'}(\omega)$ nullifies. 
One can then, for example, replace $\int  d\omega R^\alpha_\pm(\omega)\mbox{sc}\left(\frac{(\omega-\omega_{nm})\delta_0}{2}\right)\mbox{sc}\left(\frac{(\omega-\omega_{n'm'})\delta_0}{2}\right)$ by $ R^\alpha_\pm((\omega_{nm}+\omega_{n'm'})/2)$ in $\hat{M}_{\boldsymbol{\lambda}}(t+\delta,t)$ whenever $|\omega_{nm}-\omega_{n'm'}|<\delta_0^{-1}$. 
One then obtains a QME satisfying the GQDB and \eqref{eq:cond-1stlaw2}, hence the average first law \eqref{eq:1stlaw}, but not \eqref{eq:loccons}, as explained in \cite{suppmat}, section 7. 

Several approximation schemes can be applied to the Redfield QME to produce QMEs in the Lindblad form that capture coherent effects beyond the secular approximation  \cite{wacker2018, ptaszynski2019thermodynamics,farina2019, mccauley2020accurate, nathan2020universal, potts2021}. Counting statistics in these models is studied by performing the same approximation schemes on the Redfield QME with counting fields. In some cases, the resulting tilted QME falls into the class of QMEs which satisfy the GQDB and the first and second law of thermodynamics at the average level, \eqref{eq:1stlaw} and \eqref{eq:2ndlaw}, as shown in \cite{suppmat}, section 8.
An example of this is given by the symmetrized QMEs \cite{ptaszynski2019thermodynamics,mccauley2020accurate} that are in agreement with \eqref{eq:detailed} and the condition \eqref{eq:cond-1stlaw2}.
These approaches become particularly relevant in regimes where the Bohr frequencies are nearly degenerate and the secular approximation fails, as we illustrate in \cite{suppmat}, section 9, by studying heat transfers between a three level system and a bath. 

As a final remark, we note that the first and second law in non secular Lindblad QMEs are valid within second order in the interaction, which was to be expected since the starting point of these procedures (the Redfield QME) is itself obtained from a perturbative development to second order, see \cite{suppmat}, section 10. 
The second law being true only up to corrections is closely connected to the steady state of the QMEs not being of strict Gibbs form. This seemingly breaks the second law \eqref{eq:2ndlaw}, but only negligibly within the weak coupling limit.

\emph{Conclusions:}
Using counting statistics, we identified two independent conditions which ensure the thermodynamic consistency of QMEs at the fluctuating level. 
The GQDB condition \eqref{eq:detailed} guarantees the validity of the detailed FTs \eqref{eq:sym-uni-W} and \eqref{eq:sym-uni-entropy}, and thus of the average second law \eqref{eq:2ndlaw}. 
The condition \eqref{eq:loccons} guarantees a strict energy conservation at the fluctuating level and thus the average first law \eqref{eq:1stlaw}. 
Obtaining QMEs satisfying both \eqref{eq:detailed} and \eqref{eq:loccons} requires to perform the secular approximation. Beyond the secular approximation, a larger class of QMEs can satisfy FTs as well as the first and second law at the averaged level, but energy conservation at the fluctuating level is lost.

Our work suggests that the only way to achieve thermodynamic consistency at the fluctuating level beyond the secular approximation is to go beyond the semigroup hypothesis. 
In that respect, it is interesting to note that thermodynamically consistent quantum maps such as the thermalizing scattering maps recently derived in \cite{jacob2021} or the thermal operations of refs. \cite{oppenheim2013,brandao2013,ng2015}, display the same features as secular maps, in particular the decoupling of the population and coherence dynamics in the energy eigenbasis. 

We focused on time-independent system Hamiltonians, but a promising future development would be to extend our results to time-dependent ones. This should be possible 
provided that the characteristic time scale of the driving, $\tau_d$, satisfies either $\tau_d\gg\delta_0$ or $\delta_0\gg\tau_d\gg\tau_B$ where $\tau_B$ is the bath characteristic time. Indeed, in these two cases the coarse graining of the dynamics over the time scale $\delta_0$ (in the sense of \eqref{eq:teorlind}) is well defined, allowing to derive a time local QME. 
Indeed for periodically driven systems, a tilted QME can be derived in the Floquet basis, and after performing a secular approximation in that basis, the resulting QMEs has been shown to be thermodynamically consistent at the fluctuating level \cite{cuetara2015rapid}.
The optical Bloch equation has also been shown to be thermodynamically consistent at the average level \cite{elouard2020thermodynamics} but the fluctuating level remains to be investigated. The generalization of our analysis to time-dependent Hamiltonians inducing work on the system is left for future work.

This work was supported by the Luxembourg National Research Fund: A. S. by Project ThermoQO C21/MS/15713841 and V. C. by Project QUTHERM C18/MS/12704391.

\let\oldaddcontentsline\addcontentsline
\renewcommand{\addcontentsline}[3]{}
\bibliography{bibliography}
\let\addcontentsline\oldaddcontentsline


\pagebreak
\beginsupplement
\widetext
\begin{center}
\textbf{\large Supplementary material for \\
Thermodynamic consistency of quantum master equations}

\medskip

Ariane Soret$^{\dagger,*}$, Vasco Cavina$^\dagger$, Massimiliano Esposito$^\dagger$

\textit{\small{$^\dagger$Complex Systems and Statistical Mechanics, Department of Physics and Materials Science, University of Luxembourg, L-1511 Luxembourg, Luxembourg}}
\end{center}
\tableofcontents

\section{1 - Fluctuation theorems for the unitary dynamics}
\label{app:fluct1}

To prove the FT (5), we begin by using (2) and (3)
to write the generating function explicitly
\begin{equation}
G(t, \boldsymbol{\lambda}) = Tr[ \hat{U}(0,t) e^{-\lambda_S \hat{H}_S(0) - \sum_{\alpha=1}^N \lambda_\alpha \hat{H}_\alpha} \bar{\hat{\rho}}_0  \hat{U}^{\dagger}(0,t) e^{\lambda_S \hat{H}_S(t) + \sum_{\alpha=1}^N \lambda_\alpha \hat{H}_\alpha} ],    \label{eq:genapp1}
\end{equation}
while the generating function in the reverse process reads
\begin{equation}
G^R(t, \boldsymbol{\lambda}) = Tr[ \hat{U}^{\dagger}(0,t) e^{-\lambda_S \hat{H}_S(t) - \sum_{\alpha=1}^N \lambda_\alpha \hat{H}_\alpha} \bar{\hat{\rho}}_0^R  \hat{U}(0,t) e^{\lambda_S \hat{H}_S(0) + \sum_{\alpha=1}^N \lambda_\alpha \hat{H}_\alpha} ],    \label{eq:genapp2}
\end{equation}
as we can derive from (4).
In the hypotheses of the fluctuation theorem, we choose the initial state of the forward and reversed evolution to be 
\begin{equation}
 \hat{\rho}_0= \frac{e^{- \beta_S \hat{H}(0)}}{Z_S(0)} \bigotimes_\alpha \frac{e^{- \beta_\alpha \hat{H}_\alpha}}{Z_\alpha}, \quad \, \hat{\rho}_0^R=\frac{e^{-\beta_S \hat{H}_S(t)}}{Z_S(t)}\bigotimes_\alpha \frac{e^{- \beta_\alpha \hat{H}_\alpha}}{Z_\alpha}. 
 \label{eq:hyp}
\end{equation}
Within these assumptions we have 
\begin{align}
G(t, \boldsymbol{\lambda}) & = \frac{1}{Z_S(0) \prod_\alpha Z_\alpha} Tr[ \hat{U}(0,t) e^{-\lambda_S \hat{H}_S(0) - \sum_{\alpha=1}^N \lambda_\alpha \hat{H}_\alpha} e^{- \beta_S \hat{H}_S(0) - \sum_{\alpha=1}^N \beta_\alpha \hat{H}_\alpha} \hat{U}^{\dagger}(0,t) e^{\lambda_S \hat{H}_S(t) + \sum_{\alpha=1}^N \lambda_\alpha \hat{H}_\alpha} ], \\
G^R(t, \boldsymbol{\lambda}) & =\frac{1}{Z_S(t) \prod_\alpha Z_\alpha} Tr[ \hat{U}^{\dagger}(0,t) e^{-\lambda_S \hat{H}_S(t) - \sum_{\alpha=1}^N \lambda_\alpha \hat{H}_\alpha} e^{- \beta_S \hat{H}_S(t) - \sum_{\alpha=1}^N \beta_\alpha \hat{H}_\alpha} \hat{U}(0,t) e^{\lambda_S \hat{H}_S(0) + \sum_{\alpha=1}^N \lambda_\alpha \hat{H}_\alpha} ].
\end{align}
Consider now the quantity
\begin{equation}
\hspace{-0.5cm}G^R(t,-\boldsymbol{\lambda}-\boldsymbol{\beta})  =\frac{1}{Z_S(t) \prod_\alpha Z_\alpha} Tr[ \hat{U}^{\dagger}(0,t) e^{\lambda_S \hat{H}_S(t) + \sum_{\alpha=1}^N \lambda_\alpha \hat{H}_\alpha}  \hat{U}(0,t) e^{-\lambda_S \hat{H}_S(0) - \sum_{\alpha=1}^N \lambda_\alpha \hat{H}_\alpha} e^{- \beta_S \hat{H}_S(0) - \sum_{\alpha=1}^N \beta_\alpha \hat{H}_\alpha} ]. \label{eq:appgenRR}
\end{equation}
With the help of \eqref{eq:hyp} we recognize the last factor in the trace as the unnormalized initial state of the forward evolution, in such a way that
\begin{equation}
    G^R(t,-\boldsymbol{\lambda}-\boldsymbol{\beta})  =\frac{Z_S(0)}{Z_S(t)} Tr[ \hat{U}^{\dagger}(0,t) e^{\lambda_S \hat{H}_S(t) + \sum_{\alpha=1}^N \lambda_\alpha \hat{H}_\alpha}  \hat{U}(0,t) e^{-\lambda_S \hat{H}_S(0) - \sum_{\alpha=1}^N \lambda_\alpha \hat{H}_\alpha} \hat{\rho}_0]. \label{eq:appgenRR2}
\end{equation}
From \eqref{eq:appgenRR2} and \eqref{eq:genapp1}, it is immediate to find the result (5) after applying the cyclic property of trace.

We now turn to the FT (6) for the entropy. Recall that the fluctuating entropy is defined as $ \hat{S}(t)+\sum_\alpha\beta_\alpha \hat{H}_\alpha= -\log\hat{\rho}_S(t)+\sum_\alpha\beta_\alpha \hat{H}_\alpha$. The corresponding generating function is
\begin{equation}
G_{\Sigma}(t,\lambda_\Sigma)=Tr\left[ e^{-\lambda_\Sigma\log\hat{\rho}_S(t)}\prod\limits_\alpha e^{\beta_\alpha\lambda_\Sigma \hat{H}_\alpha}\hat{U}(t,0) e^{\lambda_\Sigma\log\hat{\rho}_S(0)}\prod\limits_\alpha e^{-\beta_\alpha\lambda_\Sigma \hat{H}_\alpha}\bar{\hat{\rho}}_0 \hat{U}^\dagger(t,0) \right],
\label{cgf-sigma}
\end{equation}
and the generating function for the reversed process is given by
\begin{equation}
G^R_{ \Sigma}(t,\lambda_\Sigma)=Tr\left[ e^{-\lambda_\Sigma\log\hat{\rho}_S(0)}\prod\limits_\alpha e^{\beta_\alpha\lambda_\Sigma \hat{H}_\alpha}\hat{U}^\dagger(t,0) e^{\lambda_\Sigma\log\hat{\rho}_S(t)}\prod\limits_\alpha e^{-\beta_\alpha\lambda_\Sigma \hat{H}_\alpha}\hat{\rho}_0 \hat{U}(t,0) \right].
\end{equation}
Assuming that $\hat{\rho}(0)= \hat{\rho}_S(0)\otimes\prod_\alpha e^{-\beta_\alpha \hat{H}_\alpha}/Z_\alpha$ and $\hat{\rho}^R(0)= \hat{\rho}_S(t)\otimes\prod_\alpha e^{-\beta_\alpha \hat{H}_\alpha}/Z_\alpha$, we readily check that 
\begin{equation}
    G^{R}_{\Sigma}(t,-\lambda_{\Sigma}-1)=G_{\Sigma}(t,\lambda_{\Sigma}),
    \label{eq:ft-appendix}
\end{equation}
which is the fluctuation theorem (6).

We may now easily deduce the second law of thermodynamics. Taking $\lambda_\Sigma=-1$ in \eqref{eq:ft-appendix} leads to $G_{\Delta \Sigma}(t,-1)=1$. Expressing now $G_{ \Sigma}(\lambda_{\Sigma},t)$ in terms of the probability distribution $P(\Sigma)$ of $\Sigma$ \cite{esposito2009nonequilibrium}: 
\begin{equation}
G_{ \Sigma}(\lambda_{\Sigma},t) = \int d \Sigma e^{\lambda_{\Sigma}  \Sigma}P( \Sigma)\, ,
\end{equation}
we obtain the average fluctuation theorem
\begin{equation}
\begin{array}{ll}
G_{ \Sigma}(-1,t) = \langle e^{- \Sigma}\rangle = 1 \, ,
\end{array}
\end{equation}
and, using Jensen inequality, we recover the second law of thermodynamics: $\langle  \Sigma \rangle \geq 0$.

\section{2 - Coarse graining of the unitary dynamics under the semigroup hypothesis with counting fields} \label{app:dlind}

Starting from (8) we consider the following evolution
\begin{equation}  \hat{M}_{\boldsymbol{\lambda}}(t,0)\hat{\rho}^{\boldsymbol{\lambda}}_S(0)  = \sum_{\mu, \nu} \hat{W}_{\mu,\nu}^{\boldsymbol{\lambda}} (t,0) \hat{\rho}^{\boldsymbol{\lambda}}_S(0)
\hat{W}_{\mu,\nu}^{\boldsymbol{\lambda}\, \dagger}(t,0), \label{eq:quant-map} \end{equation}
with $\hat{M}_{\boldsymbol{\lambda}}(t,0)$ denoting the dynamical map which describes the system evolution, and satisfying the semigroup hypothesis
$\hat{M}_{\boldsymbol{\lambda}}(t,0)= \hat{M}_{\boldsymbol{\lambda}}(t,s) \hat{M}_{\boldsymbol{\lambda}}(s,0)$ for every $s \in [0,t]$.
Let us now follow a strategy similar to the one presented in \cite{breuer2002theory} to derive a tilted superoperator, which assumes a Lindblad form when the counting fields are set to zero. 

First, we identify a set of operators $\hat{F}_i$, with $i=1,...N^2$, as a basis of the Hilbert space of the system, such that $Tr[\hat{F}_{i}] =0$ for $i\neq N^2$ and $\hat{F}_{N^2}= \frac{\mathds{I}}{\sqrt{N}}$.
Decomposing the $\hat{M}_{\boldsymbol{\lambda}}$ operator in this basis, we have
\begin{equation}   \hat{M}_{\boldsymbol{\lambda}}(t_1,t_2) \hat{\rho}_S(t_2) = \sum_{i,j} c_{ij}^{\boldsymbol{\lambda}}(t_2,t_1) \hat{F}_i \hat{\rho}(t_1) \hat{F}_j^{\dagger}, \quad  
c_{ij}^{\boldsymbol{\lambda}}(t_2,t_1)=  \sum_{\mu,\nu}Tr[\hat{F}_i^\dagger  \hat{W}_{\mu,\nu}^{\boldsymbol{\lambda}} (t_2,t_1)]
 Tr[\hat{F}_j^\dagger \hat{W}_{\mu,\nu}^{\boldsymbol{\lambda}} (t_2,t_1)]^*. \label{eq:defcoeff}  \end{equation}
Now we consider an infinitesimal expansion of the evolution operator
$\hat{M}_{\boldsymbol{\lambda}}(t+ \delta, t)$ such that
\begin{align}\hspace{-0.2cm}   \hat{M}_{\boldsymbol{\lambda}}(t+ \delta,t) \hat{\rho}_S =
 \frac{c_{N^2N^2}^{\boldsymbol{\lambda}}(t+ \delta,t)}{N} \hat{\rho}_S 
+ \frac{1}{\sqrt{N}} \sum_{i=1}^{N^2-1} (c_{i N^2}^{\boldsymbol{\lambda}}(t+ \delta,t) \hat{F}_i \hat{\rho}_S + c^{\boldsymbol{\lambda}}_{N^2 i}(t+ \delta,t) \hat{\rho}_S \hat{F}_i^{\dagger}
)  + \sum_{i,j=1}^{N^2-1} c_{ij}^{\boldsymbol{\lambda}}(t+ \delta,t) \hat{F}_i \hat{\rho}_S \hat{F}_j^{\dagger}. \label{eq:appclim}
\end{align}
In the limit $\delta \rightarrow \delta_0$, we obtain the tilted generator of the dynamics $\mathcal{L}_{\boldsymbol{\lambda}}(t)\hat{\rho}^{\boldsymbol{\lambda}}_S(t)  = \lim_{\delta \rightarrow \delta_0} \frac{\hat{M}_{\boldsymbol{\lambda}}(t+ \delta,t) - \mathds{I}  }{\delta} \hat{\rho}^{\boldsymbol{\lambda}}_S(t) $ as defined in (9). To obtain a generator of the form (10), we begin by 
defining
\begin{align}
\hat{F}_{\boldsymbol{\lambda}}(t) = \lim_{\delta \rightarrow \delta_0} \frac{1}{\delta}\sum_{i=1}^{N^2-1} c_{i N^2}^{\boldsymbol{\lambda}}(t+ \delta,t) \hat{F}_i; \quad  \hat{G}_{\boldsymbol{\lambda}}(t)=\frac{1}{2} ( \hat{F}_{\boldsymbol{\lambda}}(t)+ \hat{F}_{\boldsymbol{\lambda}}^\dagger(t) ) +\frac{1}{2 N} k^{\boldsymbol{\lambda}}(t); \quad H'_{\boldsymbol{\lambda}}(t) = \frac{1}{2 i}(-\hat{F}_{\boldsymbol{\lambda}}(t) +
\hat{F}_{\boldsymbol{\lambda}}^\dagger(t)), \label{eq:appFGdef}
\end{align}
where $k^{\boldsymbol{\lambda}}(t) = \lim_{\delta \rightarrow \delta_0}  \frac{c_{N^2N^2}^{\boldsymbol{\lambda}}(t+ \delta,t) -N}{\delta}$ with $k^{\boldsymbol{\lambda}}(t) = k^{\boldsymbol{\lambda}\,*}(t)$.

Using the definitions above we obtain a tilted generator
\begin{equation}  \mathcal{L}_{\boldsymbol{\lambda}}(t)\hat{\rho}^{\boldsymbol{\lambda}}_S(t)= - i [\hat{H}'_{\boldsymbol{\lambda}}(t), \hat{\rho}^{\boldsymbol{\lambda}}_S(t)] + \{\hat{G}_{\boldsymbol{\lambda}}(t), \hat{\rho}^{\boldsymbol{\lambda}}_S(t)\} + \sum_{i,j=1}^{N^2-1} a_{ij}^{\boldsymbol{\lambda}}(t) \hat{F}_i \hat{\rho}^{\boldsymbol{\lambda}}_S(t) \hat{F}_j^{\dagger}, \label{eq:gengenapp1}
\end{equation}
with $a_{ij}^{\boldsymbol{\lambda}}(t) = \lim_{\delta \rightarrow \delta_0} \frac{1}{\delta} c_{ij}^{\boldsymbol{\lambda}}(t+ \delta,t)$. Setting the system counting field to zero, $\lambda_S=0$, we obtain a tilted generator in the form (10),
\begin{equation}  
\mathcal{L}_{0,\boldsymbol{\lambda_B}}(t)\hat{\rho}^{0,\boldsymbol{\lambda_B}}_S(t)= - i [\hat{H}'_{0,\boldsymbol{\lambda_B}}(t), \hat{\rho}_S^{0,\boldsymbol{\lambda_B}}(t)] + \mathcal{D}_{0,\boldsymbol{\lambda_B}}(t)\hat{\rho}_S^{0,\boldsymbol{\lambda_B}}(t),
\label{eq:lind-app}
\end{equation}
with $\mathcal{D}_{0,\boldsymbol{\lambda_B}}(t)\hat{\rho}_S^{0,\boldsymbol{\lambda_B}}(t)=\{\hat{G}_{0,\boldsymbol{\lambda_B}}(t), \hat{\rho}^{0,\boldsymbol{\lambda_B}}_S(t)\} +\mathcal{J}_{0,\boldsymbol{\lambda}_B}(t) \hat{\rho}_S^{0,\boldsymbol{\lambda}_B}(t)$ and 
\begin{equation}
    \mathcal{J}_{0,\boldsymbol{\lambda}_B}(t) \hat{\rho}_S^{0,\boldsymbol{\lambda}_B}(t)= \sum_{i,j=1}^{N^2-1} a_{ij}^{0,\boldsymbol{\lambda_B}}(t) \hat{F}_i \hat{\rho}_S^{0,\boldsymbol{\lambda_B}}(t) \hat{F}_j^{\dagger}.
\end{equation}
When all the counting fields are set to zero, \eqref{eq:lind-app} is the so called first standard form of the generator \cite{breuer2002theory}.
Note that, since, when the counting fields are set to zero, the matrix of coefficients $[a_{ij}]$ is positive, we may diagonalize the dissipator $\mathcal{D}_{\boldsymbol{0}}(t) $ in order to obtain a QME of Lindblad form \cite{breuer2002theory}.

\medskip

We now turn to the time reversed process, and study the relation between the forward and reversed Kraus operators. The Kraus operators of the reversed process are defined as
\begin{equation}
   \hat{W}_{\mu,\nu}^{R\,\boldsymbol{\lambda}} = \sqrt{\eta_{\nu}} \langle \mu | \hat{U}_{-\boldsymbol{\lambda}}^{\dagger}(t,0)  |\nu \rangle 
\end{equation}
Assuming that the baths Hamiltonians commute, we may write $|\nu\rangle=|\nu_1\rangle\otimes...\otimes|\nu_N\rangle$, where $\{|\nu_\alpha\rangle\}$ denote the eigenstates of $\hat{H}_\alpha$, with eigenvalues $\{\nu_\alpha\}$, s.t. $\nu=\sum_\alpha\nu_\alpha$.
By explicitly expanding the definitions, we obtain
\begin{align}
 \hat{W}_{\mu,\nu}^{R\,\boldsymbol{\lambda}} & = \sqrt{\eta_{\nu}} 
 e^{\frac{\lambda_S}{2} \hat{H}_S(0)}  \langle \mu | \hat{U}^{\dagger}(t,0) | \nu \rangle  e^{-\frac{\lambda_S}{2} \hat{H}_S(t)} e^{\sum_{\alpha=1}^N \frac{\lambda_\alpha}{2} (\mu_{\alpha} -\nu_{\alpha}) }, \label{eq:appC1}\\
  \hat{W}_{\mu,\nu}^{\boldsymbol{\lambda}} & = \sqrt{\eta_{\nu}} 
 e^{\frac{\lambda_S}{2} \hat{H}_S(t)}  \langle \mu | \hat{U}(t,0) | \nu \rangle  e^{-\frac{\lambda_S}{2} \hat{H}_S(0)} e^{\sum_{\alpha=1}^N \frac{\lambda_\alpha}{2} (\mu_{\alpha} -\nu_{\alpha}) } \, .
\end{align}
We can easily verify that
\begin{align}
      (\hat{W}_{\nu,\mu}^{\boldsymbol{\lambda}})^{\dagger} & = \sqrt{\eta_{\mu}} 
 e^{-\frac{\lambda_S}{2} \hat{H}_S(0)}  \langle \nu | \hat{U}(t,0) | \mu \rangle ^{\dagger} e^{\frac{\lambda_S}{2} \hat{H}_S(t)} e^{-\sum_{\alpha=1}^N \frac{\lambda_\alpha}{2} (\mu_{\alpha } -\nu_{\alpha}) }.
\end{align}
After replacing $\sqrt{\eta_{\mu}} = \sqrt{\eta_{\nu}} e^{-\sum_{\alpha=1}^N\frac{\beta_\alpha}{2} (\mu_\alpha - \nu_\alpha)}$ and $ \langle \nu | \hat{U}(t,0) | \mu \rangle ^{\dagger} =
\langle \mu | \hat{U}^{\dagger}(t,0) | \nu \rangle $ in the equation above, we are left with 
\begin{align}
      (\hat{W}_{\nu,\mu}^{\boldsymbol{\lambda}})^{\dagger} & = \sqrt{\eta_{\nu}} 
 e^{-\frac{\lambda_S}{2} \hat{H}_S(0)}   \langle \mu | \hat{U}^{\dagger}(t,0) | \nu \rangle e^{\frac{\lambda_S}{2} \hat{H}_S(t)} e^{-\sum_{\alpha=1}^N \frac{\lambda_\alpha + \beta_\alpha}{2} (\mu_{\alpha} -\nu_{\alpha}) }. \label{eq:appC2}
\end{align}
By comparing \eqref{eq:appC1} and \eqref{eq:appC2} we finally obtain (12)
\begin{equation}
  \hat{W}_{\mu,\nu}^{R\,\boldsymbol{\lambda}}=(\hat{W}_{\nu,\mu}^{(-\lambda_S, -\boldsymbol{\lambda}_B -\boldsymbol{\beta}_B)} )^{\dagger} =e^{\lambda_S \hat{H}_S(0) }(\hat{W}_{\nu,\mu}^{(\lambda_S, -\boldsymbol{\lambda}_B -\boldsymbol{\beta}_B)} )^{\dagger} e^{-\lambda_S \hat{H}_S(t) }.
\end{equation}
\section{3 - Quantum fluctuation theorems for QMEs}\label{app:ft-q}
In order to prove that the FTs (5) and (6) hold for open systems under the GQDB condition (13), we begin by noticing that the property $\hat{W}_{\mu\nu}^{\boldsymbol{\lambda}}(t,0)=e^{\lambda_S\hat{H}_S(t)/2}\hat{W}_{\mu\nu}^{0,\boldsymbol{\lambda_B}}(t,0)e^{-\lambda_S\hat{H}_S(0)/2}$ of the Kraus operators
results, by replacing in (8), in a tilted density matrix of the form 
\begin{equation}
\begin{array}{l}
    \hat{\rho}_S^{\boldsymbol{\lambda}}(t) =e^{\frac{\lambda_S}{2} \hat{H}_S(t)} \mathcal{T}\big\{e^{\int_0^t \mathcal{L}_{0,\boldsymbol{\lambda}_B}(\tau)d \tau }\big\}[e^{-\frac{\lambda_S}{2} \hat{H}_S(0)} \bar{\hat{\rho}}_S (0)  e^{-\frac{\lambda_S}{2} \hat{H}_S(0)}] e^{\frac{\lambda_S}{2} \hat{H}_S(t)}.
\end{array}
\label{eq:gen-tot}
\end{equation} 

We start by proving (5). We assume that $\hat{\rho}_S(0)=e^{-\beta_S \hat{H}_S(0)}/ Z_S(0)$ and $\hat{\rho}_S^R(0) = e^{-\beta_S \hat{H}_S(t)}/ Z_S(t)$. Then, using \eqref{eq:gen-tot} we have
\begin{equation}
\begin{array}{ll}
Tr[e^{\lambda_S \hat{H}_S(t)}e^{t\mathcal{L}_{0,\boldsymbol{\lambda_B}}} 
[e^{-\frac{\lambda_S}{2} \hat{H}_S(0)} \hat{\rho}_S(0) e^{-\frac{\lambda_S}{2} \hat{H}_S(0)}] ] &=Tr[(e^{t\mathcal{L}^\dagger_{0,\boldsymbol{\lambda_B}}}[e^{\lambda_S \hat{H}_S(t)}])^\dagger e^{-\frac{\lambda_S}{2} \hat{H}_S(0)} \hat{\rho}_S(0) e^{-\frac{\lambda_S}{2} \hat{H}_S(0)}]\\
\\
&= Tr[(e^{t\mathcal{L}^R_{0,\boldsymbol{-\lambda_B-\beta_B}}} [e^{\lambda_S \hat{H}_S(t)}])^\dagger e^{-(\lambda_S +\beta_S)\hat{H}_S(0)} ]/Z_S(0),
\end{array}
\end{equation}
Since $e^{t\mathcal{L}^R_{0,\boldsymbol{-\lambda_B-\beta_B}}} [e^{\lambda_S \hat{H}_S}]$ is self-adjoint, we can remove the dagger in the second line and obtain
\begin{equation}
    \begin{array}{l}
Tr[(e^{t\mathcal{L}^R_{0,\boldsymbol{-\lambda_B-\beta_B}}} [e^{\lambda_S \hat{H}_S(t)}])^\dagger e^{-(\lambda_S +\beta_S)\hat{H}_S(0)} ]/Z_S(0) = Tr_S[e^{t\mathcal{L}^R_{0,\boldsymbol{-\lambda_B-\beta_B}}} [e^{\lambda_S \hat{H}_S(t)}] e^{-(\lambda_S +\beta_S)\hat{H}_S(0)} ]/Z_S(0)\\
\\
= \frac{Z_S(t)}{Z_S(0)} Tr_S[e^{t\mathcal{L}^R_{0,\boldsymbol{-\lambda_B-\beta_B}}}[\hat{\rho}_S^R(0) e^{(\lambda_S+ \beta_S) \hat{H}_S(t)} ]e^{-(\lambda_S +\beta_S)\hat{H}_S(0)} ].
    \end{array}
    \label{ft-lin-aux-2}
\end{equation}
Thus, we have 
\begin{equation}
    Tr[\hat{\rho}_{S,\boldsymbol{\lambda}}(t)]=Tr[\hat{\rho}_{S,\boldsymbol{-\lambda-\beta}}^R(t)]\frac{Z_S(t)}{Z_S(0)},
\end{equation}
which completes the proof.

The proof of the FT for the entropy production follows the same steps. This time, the only assumption required is $\hat{\rho}_S(0)=\hat{\rho}_S^R(t)$ and $\hat{\rho}^R_S(0)=\hat{\rho}_S(t)$. The MGF is given by
\begin{equation}
\begin{array}{ll}
Tr_S[\hat{\rho}_{S}^{\lambda_\Sigma}(t)]&=Tr[e^{-\lambda_\Sigma \log\hat{\rho}_S(t)}e^{t\mathcal{L}_{0,\lambda_\Sigma\boldsymbol{\beta_B}}} 
[e^{\frac{\lambda_\Sigma}{2} \log\hat{\rho}_S(0)} \hat{\rho}_S(0) e^{\frac{\lambda_\Sigma}{2} \log\hat{\rho}_S(0)} ]]\\
\\
&=Tr[(e^{t\mathcal{L}^\dagger_{0,\lambda_\Sigma\boldsymbol{\beta_B}}}[e^{-\lambda_\Sigma \log\hat{\rho}_S(t)}])^\dagger e^{\lambda_\Sigma\log\hat{\rho}_S(0)} \hat{\rho}_S(0)]\\
\\
&= Tr[(e^{t\mathcal{L}^R_{0,-(\lambda_\Sigma+1)\boldsymbol{\beta_B}}} [e^{-\lambda_\Sigma \log\hat{\rho}_S(t)}])^\dagger e^{\lambda_\Sigma\log\hat{\rho}_S(0)} \hat{\rho}_S(0) ]\\
\\
&= Tr[(e^{t\mathcal{L}^R_{0,-(\lambda_\Sigma+1)\boldsymbol{\beta_B}}} [e^{-(\lambda_\Sigma+1) \log\hat{\rho}_S(t)}\hat{\rho}_S(t)])^\dagger e^{(\lambda_\Sigma+1)\log\hat{\rho}_S(0)} ]\\
\\
&=Tr_S[\hat{\rho}^R_{S,-\lambda_\Sigma-1}(t)]\, ,
\end{array}
\end{equation}
which completes the proof.

\section{4 - Relaxation to a Gibbs state}\label{app:commute}

The relation between the forward and time reversed generators $\mathcal{L}$ and $\mathcal{L}^R$ is obtained from the unitary dynamics, where the time reversed process is given by \cite{esposito2009nonequilibrium}
\begin{equation}
    \hat{\rho}^R(t)=\hat{U}^\dagger(t,0)\hat{\rho}^R(0)\hat{U}(t,0), \label{eq:unitapp}
\end{equation}
from which we obtain the equation
\begin{equation}
    \frac{d\hat{\rho}^R(t)}{dt}=i[\hat{H}(t),\hat{\rho}^R(t)].
    \label{eq:eq-tr}
\end{equation}
To derive the equation above is not necessary to assume that the Hamiltonian is time reversal invariant.
Indeed, as discussed in \cite{esposito2009nonequilibrium}, the property \eqref{eq:unitapp} is still valid even in presence of non invariant components of $\hat{H}$, provided that we suitably invert these components when discussing time reversed processes. 
Let us use the short hand notation $\hat{\boldsymbol{B}}$ to denote the non time reversal invariant components of $\hat{H}$, e.g., magnetic fields or spins. Writing the dependence in $\hat{\boldsymbol{B}}$ explicitly in the Hamiltonian, $\hat{H}(\hat{\boldsymbol{B}})$, we have the relation
\begin{equation}
    \Theta\hat{H}(\hat{\boldsymbol{B}},t)\Theta=\hat{H}(-\hat{\boldsymbol{B}},t) \, .
\end{equation}
Using the properties $\Theta^2=\mathds{I}$, $\Theta i \Theta = -i$ \cite{esposito2009nonequilibrium} and limiting ourselves to the time independent case $\hat{H}(\boldsymbol{B},t) = \hat{H}(\boldsymbol{B})$, we may re-write \eqref{eq:eq-tr} as
\begin{equation}
    \frac{d\hat{\rho}^R(t)}{dt} = \Theta (-i[\Theta\hat{H}(\hat{\boldsymbol{B}})\Theta,\Theta\hat{\rho}^R(t)\Theta])\Theta = \Theta (-i[\hat{H}(-\hat{\boldsymbol{B}}),\Theta\hat{\rho}^R(t)\Theta])\Theta \, ,
\end{equation}
from which, after tracing out the degrees of freedom of the bath, we obtain
\begin{equation}
    \mathcal{L}^R_{\hat{\boldsymbol{B}}}(...) = \Theta\mathcal{L}_{-\hat{\boldsymbol{B}}}(\Theta...\Theta)\Theta \, ,
    \label{eq:fwd-r}
\end{equation}
where the subscript $\hat{\boldsymbol{B}}$ allows to keep track of the dependence of the generator in $\hat{\boldsymbol{B}}$. It is then immediate to check that, if the Gibbs state is a fixed point of $\mathcal{L}_{\hat{\boldsymbol{B}}}^R$, then it is also a fixed point of $\mathcal{L}_{\hat{\boldsymbol{B}}}$, using that, under time reversal, the Gibbs state becomes 
\begin{equation}
    \Theta \frac{e^{-\beta\hat{H}_S(\hat{\boldsymbol{B}})}}{Z_S}\Theta=\frac{e^{-\beta\hat{H}_S(-\hat{\boldsymbol{B}})}}{Z_S},
\end{equation}
with $Z_S=Tr_S[e^{-\beta\hat{H}_S(\hat{\boldsymbol{B}})}]$.

\section{5 - Alternative definition of detailed balance}
\label{app:neweq}

An alternative definition of detailed balance, often used in the literature \cite{fagnola2007generators,fagnola2008detailed}, is given by
\begin{equation}
   \mathcal{L}^{R}[...] = \sqrt{\hat{\rho}_{ss}} \mathcal{L}^{\dag}[\sqrt{\hat{\rho}_{ss}^{-1}} ...  \sqrt{\hat{\rho}_{ss}^{-1}}] \sqrt{\hat{\rho}_{ss}}, \label{eq:det2}
\end{equation}
where $\hat{\rho}_{ss}$ is a fixed point of the dynamics, $\mathcal{L}(\hat{\rho}_{ss})=0$.
The detailed balance condition \eqref{eq:det2} can formally be interpreted as imposing the following translation symmetry, 
\begin{equation} \mathcal{L}^{( \Sigma_{ss})}_{\lambda_{ss}}[...] =  
\mathcal{L}^{( \Sigma_{ss})}_{\lambda_{ss}+ \chi}[...], \label{eq:locinv}\end{equation}
where $\mathcal{L}^{( \Sigma_{ss})}_{\lambda_{ss}}$ is the superoperator obtained by putting a counting field on $\hat{\Sigma}_{ss} = - \log(\hat{\rho}_{ss}) + \sum_{\alpha} \beta_{\alpha} \hat{H}_{\alpha}$. Hence
\eqref{eq:locinv} is different from (15). 

To see the connection between \eqref{eq:det2} and \eqref{eq:locinv}, we start from (13) with $\boldsymbol{\lambda_B}=\boldsymbol{0}$, and use the fact that 
$$\mathcal{L}_{\lambda_{ss}}^{(\Sigma_{ss})} = e^{-\lambda_{ss}\log\hat{\rho}_{ss}/2}\mathcal{L}_{0,\lambda_{ss}\boldsymbol{\beta_B}}[e^{\lambda_{ss}\log\hat{\rho}_{ss}/2}\hat{\bar{\rho}}_0e ^{\lambda_{ss}\log\hat{\rho}_{ss}/2}]e^{-\lambda_{ss}\log\hat{\rho}_{ss}/2}\, .$$ 
Setting $\lambda_{ss}=1$, we obtain
\begin{equation}
\begin{array}{ll}
\mathcal{L}_{0,\boldsymbol{0}}^R[...]&=\mathcal{L}^\dagger_{0,\boldsymbol{\beta_B}}[...]=\sqrt{\hat{\rho}_{ss}}\mathcal{L}^{(\Sigma_{ss})\dagger}_{\lambda_{ss}=1}[\sqrt{\hat{\rho}_{ss}}^{-1}...\sqrt{\hat{\rho}_{ss}}^{-1}] \sqrt{\hat{\rho}_{ss}}.
\end{array}
\end{equation}
Finally, using \eqref{eq:locinv}, we may replace $\mathcal{L}^{(\Sigma_{ss})\dagger}_{\lambda_{ss}=1}=\mathcal{L}^{(\Sigma_{ss})\dagger}_{\lambda_{ss}=0}=\mathcal{L}^\dagger$, and we obtain \eqref{eq:det2}.

\section{6 - Redfield ME for counting statistics}\label{app:red}

To derive the Redfield ME with counting fields on the bath, we adapt the textbook derivation \cite{breuer2002theory}. For simplicity, we assume that $\hat{H}$ is time independent. Starting from 
\begin{equation}
\hat{\rho}^{0,\boldsymbol{\lambda_B}}(t)=\hat{U}_{0,\boldsymbol{\lambda_B}}(t,0)\hat{\rho}(0) \hat{U}^{\dagger}_{0,\boldsymbol{\lambda_B}}(t,0),
\end{equation}
we go to the interaction picture $\hat{\rho}_I^{0,\boldsymbol{\lambda_B}}(t)=\hat{U}_0(t,0)^\dagger \hat{\rho}^{0,\boldsymbol{\lambda_B}}(t)\hat{U}_0(t,0)$, with 
\begin{equation}
    \hat{U}_0(t,0)=e^{-i\hat{H}_0 t}; \quad     \hat{H}_0=\hat{H}_S+\sum\limits_{\alpha=1}^N \hat{H}_\alpha,
\end{equation}
which leads to 
\begin{equation}
    \frac{d\hat{\rho}_I^{0,\boldsymbol{\lambda_B}}(t)}{dt} = -i \hat{V}^{0,\boldsymbol{\lambda_B}}(t)\hat{\rho}_I^{0,\boldsymbol{\lambda_B}}(t)+i\hat{\rho}_I^{0,\boldsymbol{\lambda_B}}(t) \hat{V}^{0,\boldsymbol{\lambda_B}}(t),
    \label{eq:red-aux}
\end{equation}
where
\begin{equation}
\begin{array}{ll}
\hat{V}^{0,\boldsymbol{\lambda_B}}(t)&=-\hat{H}_0 + \hat{U}_0^\dagger(t,0) e^{\sum_\alpha\lambda_\alpha \hat{H}_\alpha/2} \hat{H}(t) e^{-\sum_\alpha\lambda_\alpha \hat{H}_\alpha/2} \hat{U}_0(t,0) \\
\\
&=  \sum_{\alpha'} \hat{U}^{\dagger}_0(t,0) e^{\sum_\alpha \lambda_\alpha \hat{H}_\alpha /2}  \hat{V}_{\alpha'}(t) 
     e^{-\sum_\alpha \lambda_\alpha \hat{H}_\alpha /2}  \hat{U}_0(t,0).
\end{array}
     \label{eq:defdelta}
\end{equation}
with $\hat{H}(t)$ given by (1).
Integrating the equation \eqref{eq:red-aux} gives an integral form for $\hat{\rho}_I^{0,\boldsymbol{\lambda_B}}(t)$, which we then re-inject in \eqref{eq:red-aux}. Assuming that $\hat{\rho}(t)=\hat{\rho}_S(t)\otimes\hat{\rho}_B$ at all times, and tracing out the bath degrees of freedom, we obtain, still in the interaction picture, the following equation for the system density matrix $\hat{\rho}^{0,\boldsymbol{\lambda_B}}_{S,I}$,
\begin{equation}
    \begin{array}{ll}
\frac{d\hat{\rho}_{S,I}^{0,\boldsymbol{\lambda_B}}(t)}{dt} = 
&-\int_0^t ds\, Tr_B[ \hat{V}^{0,\boldsymbol{\lambda_B}}(t) \hat{V}^{0,\boldsymbol{\lambda_B}}(s)\hat{\rho}^{0,\boldsymbol{\lambda_B}}_{S,I}(s)\otimes\hat{\rho}_B] + Tr_B[\hat{\rho}^{0,\boldsymbol{\lambda_B}}_{S,I}(s)\otimes\hat{\rho}_B \hat{V}^{0,\boldsymbol{\lambda_B}\, \dagger}(s) \hat{V}^{0,\boldsymbol{\lambda_B}\, \dagger}(t)]\\
\\
&+\int_0^t ds Tr_B[ \hat{V}^{0,\boldsymbol{\lambda_B}}(t)\hat{\rho}^{0,\boldsymbol{\lambda_B}}_{S,I}(s)\otimes\hat{\rho}_B \hat{V}^{0,\boldsymbol{\lambda_B}\, \dagger}(s)] + Tr_B[ \hat{V}^{0,\boldsymbol{\lambda_B}}(s)\hat{\rho}^{0,\boldsymbol{\lambda_B}}_{S,I}(s)\otimes\hat{\rho}_B \hat{V}^{0,\boldsymbol{\lambda_B\, \dagger}}(t)],
    \end{array}
\end{equation}
where we assumed that $Tr_B([\hat{V}^{\boldsymbol{\lambda_B}}(t),\hat{\rho}(0)])=0$. We then perform the Markov approximation, which amounts to replacing $\hat{\rho}_{S,I}(s)$ by $\hat{\rho}_{S,I}(t)$ in the integral above. Finally, we do a change of variable $s=t-s$ and take the upper bound of the integral to infinity in order to obtain a time-local markovian QME with counting fields:
\begin{equation}
    \begin{array}{l}
\frac{d\hat{\rho}_{S,I}^{0,\boldsymbol{\lambda_B}}(t)}{dt} =\\
\\

-\int_0^{+\infty} ds Tr_B[ \hat{V}^{0,\boldsymbol{\lambda_B}}(t) \hat{V}^{0,\boldsymbol{\lambda_B}}(t-s)\hat{\rho}^{0,\boldsymbol{\lambda_B}}_{S,I}(t)\otimes\hat{\rho}_B]+ Tr_B[\hat{\rho}^{0,\boldsymbol{\lambda_B}}_{S,I}(t)\otimes\hat{\rho}_B\hat{V}^{0,\boldsymbol{\lambda_B}\, \dagger}(t-s)\hat{V}^{0,\boldsymbol{\lambda_B}\, \dagger}(t)]\\
\\
+\int_0^{+\infty} ds Tr_B[\hat{V}^{0,\boldsymbol{\lambda_B}}(t)\hat{\rho}^{0,\boldsymbol{\lambda_B}}_{S,I}(t)\otimes\hat{\rho}_B\hat{V}^{0,\boldsymbol{\lambda_B}\, \dagger}(t-s)] + Tr_B[\hat{V}^{0,\boldsymbol{\lambda_B}}(t-s)\hat{\rho}^{0,\boldsymbol{\lambda_B}}_{S,I}(t)\otimes\hat{\rho}_B\hat{V}^{0,\boldsymbol{\lambda_B}\, \dagger}(t)].
    \end{array}
    \label{eq:red-aux2}
\end{equation}
Notice that setting $\boldsymbol{\lambda_B=0}$ in the equation above gives the standard Redfield equation.
To pursue the derivation, we write the interaction Hamiltonians in the general form $\hat{V}_\alpha=\hat{A}\otimes\hat{B}^{\dagger}_\alpha+\hat{A}^{\dagger}\otimes\hat{B}_\alpha$, where $\hat{A},\hat{B}_{\alpha}$ act on the system and bath Hilbert spaces respectively. We then introduce the operators
\begin{equation}
    \hat{A}_{mn}=\sum\limits_{E_{m}-E_{n}=\omega_{mn}}\Pi(E_{m})\hat{A}\Pi(E_{n}),
    \label{eq:jump-syst}
\end{equation}
where $\Pi(E)$ denotes the projector on the eigenspace of eigenvalue $E$ of the system. The operators $\hat{A}_{mn}$ satisfy the relation $[\hat{H}_S,\hat{A}_{mn}]=-\omega_{mn} \hat{A}_{mn}$, which implies $e^{i\hat{H}_St}\hat{A}_{mn} e^{-i\hat{H}_S t} = e^{-i\omega_{mn} t}\hat{A}_{mn}$. To make the connection with the notation of the main text, note that when the eigenspace of the eigenvalue $\{E_n\}$ is one dimensional, we may write $P(E_n)=|E_n\rangle\langle E_n|$, where $\{|E_n\rangle\}$ is the eigenstate of eigenvalue $E_n$. If the eigenspace is of dimension $d>1$, we may write $P(E_n)=\sum_{j=1}^{d}|E_n^{(j)}\rangle\langle E_n^{(j)}|$ where $\{E_n^{(j)}\}$ form an orthonormal basis of the eigenspace. Up to a relabelling, and allowing different operators $\hat{A}_{mn}$ to have the same $\omega_{mn}$, we can then write the operators $\hat{A}_{mn}$ in the form $\hat{A}_{mn}=g_{mn}|E_n\rangle\langle E_m|$, where $g_{mn}\in\mathbb{C}$ is a complex amplitude, $\omega_{mn}=E_{m}-E_{n}$, and where we introduced the jump operators $\hat{\sigma}_{mn}=|E_n\rangle\langle E_m|$, where $\{|E_n\rangle\}$ form an eigenbasis of $\hat{H}_S$.

Similarly, if the bath is composed by a collection of harmonic oscillators linearly coupled with the system, we may decompose $\hat{B}_\alpha=\sum_k j_{\alpha,k} \hat{b}_\alpha(\omega_{\alpha,k})$, s.t. $[\hat{H}_B,\hat{b}_\alpha(\omega_{\alpha,k})]=-\omega_{\alpha,k}\hat{b}_\alpha(\omega_{\alpha,k})$ and $[\hat{H}_\alpha,\hat{b}_\alpha^\dagger(\omega_{\alpha,k})]=\omega_{\alpha,k}\hat{b}_\alpha^\dagger(\omega_{\alpha,k})$ with $j_{\alpha,k}\in\mathbb{C}$. 

With counting fields $\boldsymbol{\lambda_B}$ on the baths, these operators become 
\begin{equation}
    \hat{V}_\alpha(\lambda_\alpha)=\hat{A}\otimes\hat{B}_\alpha^\dagger(-\lambda_\alpha)+\hat{A}^\dagger\otimes\hat{B}_\alpha(\lambda_\alpha),
    \label{eq:v-tilted}
\end{equation}
with 
\begin{equation}
\begin{array}{ll}
\hat{B}_\alpha(\lambda_\alpha) &= e^{\lambda_\alpha \hat{H}_\alpha/2}\hat{B}_\alpha e^{-\lambda_\alpha \hat{H}_\alpha/2}=\sum_k j_{\alpha,k} \hat{b}_\alpha(\omega_k)e^{-\lambda_\alpha\omega_{\alpha,k}},\\
\\
\hat{B}^\dagger_\alpha(-\lambda_\alpha) &= e^{\lambda_\alpha \hat{H}_\alpha/2}\hat{B}^\dagger_\alpha e^{-\lambda_\alpha \hat{H}_\alpha/2}=\sum_k j_{\alpha,k} \hat{b}^\dagger_\alpha(\omega_k)e^{\lambda_\alpha\omega_{\alpha,k}}.
\end{array}
  \label{eq:tilted-bath}  
\end{equation}
Replacing in \eqref{eq:red-aux2}, and going back to the Schrodinger picture, we obtain
\begin{equation}
    \frac{d\hat{\rho}_S^{0,\boldsymbol{\lambda_B}}(t)}{dt} =  -i[\hat{H}_S + \hat{H}_{LS},\hat{\rho}^{0,\boldsymbol{\lambda_B}}_S(t)]+\mathcal{D}^{Red}_{0,\boldsymbol{\lambda_B}}(t)\hat{\rho}_S^{0,\boldsymbol{\lambda_B}}(t),
    \label{eq:red-lambda}
\end{equation}
where in the equation above $\hat{H}_{LS}$ and $\mathcal{D}^{Red}_{0,\boldsymbol{\lambda}_B}$ denote the Lamb shift and the tilted dissipator, which respectively read
\begin{equation}
\hat{H}_{LS}= \sum\limits_{mn,m'n',\alpha} \big[ I^\alpha_-(\omega_{m'n'},0)\hat{A}_{mn}\hat{A}^\dagger_{m'n'} + I^\alpha_+(\omega_{m'n'},0)\hat{A}_{mn}^\dagger \hat{A}_{m'n'} \big],
\label{eq:LS-red}
\end{equation}
and
\begin{equation}
\begin{array}{ll}
\mathcal{D}^{Red}_{0,\boldsymbol{\lambda_B}}(t)\hat{\rho}_S^{0,\boldsymbol{\lambda_B}}(t) =& - \sum\limits_{mn,m'n',\alpha}  R^\alpha_-(\omega_{m'n'},0)\hat{A}_{mn}\hat{A}^\dagger_{m'n'}\hat{\rho}^{0,\boldsymbol{\lambda_B}}_S+R^\alpha_-(\omega_{mn},0)\hat{\rho}^{0,\boldsymbol{\lambda_B}}_S\hat{A}_{mn}\hat{A}^\dagger_{m'n'}\\
\\
& - \sum\limits_{mn,m'n',\alpha} R^\alpha_+(\omega_{m'n'},0)\hat{A}_{mn}^\dagger \hat{A}_{m'n'}\hat{\rho}^{0,\boldsymbol{\lambda_B}}_S+R^\alpha_+(\omega_{mn},0)\hat{\rho}^{0,\boldsymbol{\lambda_B}}_S \hat{A}^\dagger_{mn} \hat{A}_{m'n'}\\
\\
&+ \sum\limits_{mn,m'n',\alpha}(\Gamma^\alpha_+(\omega_{m'n'},\lambda_\alpha)^*+\Gamma^\alpha_+(\omega_{mn},\lambda_\alpha))\hat{A}_{mn}\hat{\rho}^{0,\boldsymbol{\lambda_B}}_S \hat{A}_{m'n'}^\dagger\\
\\
&+\sum\limits_{mn,m'n',\alpha}(\Gamma^\alpha_-(\omega_{m'n'},\lambda_\alpha)^*+\Gamma^\alpha_-(\omega_{mn},\lambda_\alpha))\hat{A}^\dagger_{mn}\hat{\rho}^{0,\boldsymbol{\lambda_B}}_S \hat{A}_{m'n'},
    \end{array}
\end{equation}
with $R^\alpha_\pm$ and $I^\alpha_\pm$ being respectively the real and imaginary parts of $\Gamma^\alpha_{\pm}(\omega_{mn},\lambda_\alpha)$, defined as
\begin{equation}
    \begin{array}{ll}
\Gamma^\alpha_+(\omega_{mn},\lambda_\alpha)=& \int_0^{+\infty}ds\, Tr_{\alpha}[\hat{B}(s,-\lambda_\alpha)\hat{B}^\dagger(0,-\lambda_\alpha)\hat{\rho}_{\alpha}]e^{i\omega_{mn} s}, \\
\\
\Gamma^\alpha_-(\omega_{mn},\lambda_\alpha)=& \int_0^{+\infty}ds\, Tr_{\alpha}[\hat{B}^\dagger(s,\lambda_\alpha)\hat{B}(0,\lambda_\alpha)\hat{\rho}_{\alpha}]e^{-i\omega_{mn} s},
    \end{array}
    \label{eq:coefs-red}
\end{equation}
and $Tr_\alpha[...]$ denotes the partial trace over the bath $\alpha$.
Since the baths are initialized in the Gibbs state, the coefficients $\Gamma_\pm^\alpha$ satisfy the symmetry
\begin{equation}
    \Gamma^\alpha_\pm(\omega_{mn},-\lambda_{\alpha}-\beta_{\alpha})=\Gamma^\alpha_\mp(\omega_{mn},\lambda_{\alpha})^* \, .
    \label{eq:sym-gamma}
\end{equation}
Following the same method, we obtain for the reversed process
\begin{equation}
\begin{array}{rl}
\frac{d\hat{\rho}^R_S(t)}{dt} =& \mathcal{L}^{Red,R}_{0,\boldsymbol{\lambda_B}}(t)\hat{\rho}_S^R,
\end{array}
\label{ft-red-1}
\end{equation}
where, to alleviate the notations, we dropped the subscript ${0,\boldsymbol{\lambda_B}}$, and where
\begin{equation}
    \mathcal{L}^{Red,R}_{0,\boldsymbol{\lambda_B}}(...)=i[\hat{H}_S+\hat{H}_{LS},...] +\mathcal{D}^{Red }_{0,\boldsymbol{\lambda_B}}(...)^*.
\end{equation}
One can readily check that the above operator is not equal to $i[\hat{H}_S+\hat{H}_{LS},...] +\mathcal{D}^{Red,\dagger }_{0,\boldsymbol{-\lambda_B-\beta_B}}(...)$. 

\section{7 - Consistent perturbative derivation of a QME} \label{app:conslim}

We provide here details on the derivation of (23), and give an example of a QME satisfying the GQDB (13) and the average energy conservation (18), but not the strict energy conservation (15). 
For ease of calculation but without loss of generality, we consider the case of a single bath, so that the counting field has two components $\boldsymbol{\lambda}=(\lambda_S,\lambda_B)$.
Recall that, in the interaction picture, the Kraus operators are of the form 
\begin{equation}
    \hat{W}^{\boldsymbol{\lambda}}_{\mu,\nu}(t,0)=\sqrt{\eta_\nu}\langle\mu | \mathcal{T}\{e^{-i\int_0^t ds \hat{V}^{\boldsymbol{\lambda}}(s)}\}|\nu\rangle = \sqrt{\eta_\nu}e^{\lambda(\mu-\nu)/2}e^{\lambda \hat{H}_S/2}\langle\mu | \mathcal{T}\{e^{-i\int_0^t ds \hat{V}(s)}\}|\nu\rangle e^{-\lambda \hat{H}_S/2},
\end{equation}
where here we have chosen $\boldsymbol{\lambda}=(\lambda,\lambda)$. 
We now decompose the operators $\hat{W}^{\boldsymbol{\lambda}}_{\mu,\nu}$ in the basis formed by jump operators $\hat{\sigma}_{mn}$ used in the main text and introduced in the section 6 of this supplemental material. Recall that different $\hat{\sigma}_{mn}$ can have the same $\omega_{mn}$. We then perform the semigroup hypothesis to obtain the superoperator,
\begin{equation}
    \mathcal{L}_{\boldsymbol{\lambda}}(t)\hat{\rho}^{\boldsymbol{\lambda}}_S(t)=\lim_{\delta\to \delta_0}\frac{1}{\delta}\left(\sum\limits_{mn,m'n'}d^{\boldsymbol{\lambda}}_{mn,m'n'}(t)\hat{\sigma}_{mn}[...]\hat{\sigma}_{m'n'}^\dagger- \mathds{1}\right)\hat{\rho}^{\boldsymbol{\lambda}}_S(t),
    \label{eq:aux-lfl}
\end{equation}
where 
\begin{equation}
\begin{array}{ll}
d^{\boldsymbol{\lambda}}_{mn,m'n'}(t)&=\sum\limits_{\mu,\nu}\eta_\nu Tr_S[\hat{\sigma}_{mn}^\dagger\hat{W}^{\boldsymbol{\lambda}}_{\mu,\nu}(t+\delta,t)]Tr_S[\hat{\sigma}_{m'n'}\hat{W}_{\mu,\nu}^{\boldsymbol{\lambda}\dagger}(t+\delta,t)]\\
\\
&=\sum\limits_{\mu,\nu}\eta_\nu \langle E_{n},\mu|\mathcal{T}_{\leftarrow}\{e^{-i\int_t^{t+\delta} ds \hat{V}^{\boldsymbol{\lambda}}(s)}\}|E_{m},\nu\rangle \langle E_{m'},\nu|\mathcal{T}_{\rightarrow}\{e^{i\int_t^{t+\delta} ds \hat{V}^{\boldsymbol{\lambda}\dagger}(s)}\}|E_{n'},\mu\rangle.
\end{array}
\label{eq:dij}
\end{equation}
We now perform a perturbative expansion to second order in $\hat{V}$, 
\begin{equation}
\hspace{-1cm}\begin{array}{ll}
    \langle E_{n},\mu|\mathcal{T}_{\leftarrow}\{e^{-i\int_t^{t+\delta} ds \hat{V}^{\boldsymbol{\lambda}}(s)}\}|E_{m},\nu\rangle &=\langle E_{n},\mu|\mathds{I} -i\int_t^{t+\delta} ds \hat{V}^{\boldsymbol{\lambda}}(s) - \frac{1}{2}\int_t^{t+\delta} ds\int_t^{s} ds' \hat{V}^{\boldsymbol{\lambda}}(s)\hat{V}^{\boldsymbol{\lambda}}(s') |E_{m},\nu\rangle + \mathcal{O}(\hat{V}^3),\\
    \\
    \langle E_{n},\mu|\mathcal{T}_{\rightarrow}\{e^{i\int_t^{t+\delta} ds \hat{V}^{\boldsymbol{\lambda}\dagger}(s)}\}|E_{m},\nu\rangle &=\langle E_{n},\mu|\mathds{I} +i\int_t^{t+\delta} ds \hat{V}^{\boldsymbol{\lambda}\dagger}(s) - \frac{1}{2}\int_t^{t+\delta} ds\int_s^{t+\delta} ds' \hat{V}^{\boldsymbol{\lambda}\dagger}(s)\hat{V}^{\boldsymbol{\lambda}\dagger}(s') |E_{m},\nu\rangle + \mathcal{O}(\hat{V}^3).
\end{array}
\end{equation}
Re-injecting in \eqref{eq:aux-lfl}, and recalling that the bath density matrix is $\hat{\rho}_B=\sum_\nu \eta_\nu |\nu\rangle\langle\nu|$ we obtain,
\begin{equation}
    \begin{array}{l}
\mathcal{L}_{\boldsymbol{\lambda}}(t)\hat{\rho}^{\boldsymbol{\lambda}}_S(t) = \lim_{\delta\to \delta_0}\frac{1}{\delta}\left(\sum\limits_{n,n'} \hat{\sigma}_{nn}\hat{\rho}_S(t)\hat{\sigma}_{n'n'}^\dagger- \hat{\rho}^{\boldsymbol{\lambda}}_S(t)\right)\\
    \\
    +\lim_{\delta\to \delta_0}\frac{1}{\delta}\sum\limits_{mn,m'n'}Tr [\int_t^{t+\delta} ds\, \hat{\sigma}_{mn}^\dagger\hat{V}^{\boldsymbol{\lambda}}(s) \hat{\rho}_B\int_t^{t+\delta} ds\, \hat{\sigma}_{m'n'}\hat{V}^{\boldsymbol{\lambda}\dagger}(s) ]\hat{\sigma}_{mn}\hat{\rho}^{\boldsymbol{\lambda}}_S(t)\hat{\sigma}_{m'n'}^\dagger\\
    \\
    -\frac{1}{2}\lim_{\delta\to \delta_0}\frac{1}{\delta}\sum\limits_{mn,m'n'}Tr_S[\hat{\sigma}_{m'n'}]Tr [\hat{\sigma}_{mn}^\dagger\int_t^{t+\delta} ds \hat{V}^{\boldsymbol{\lambda}}(s) \int_t^{s} ds' \hat{V}^{\boldsymbol{\lambda}}(s') \hat{\rho}_B]\hat{\sigma}_{mn}\hat{\rho}^{\boldsymbol{\lambda}}_S(t)\hat{\sigma}_{m'n'}^\dagger\\
    \\
    -\frac{1}{2}\lim_{\delta\to \delta_0}\frac{1}{\delta}\sum\limits_{mn,m'n'}Tr_S[\hat{\sigma}_{mn}^\dagger]Tr [\hat{\sigma}_{m'n'}\int_t^{t+\delta} ds \hat{V}^{\boldsymbol{\lambda}\dagger}(s) \int_s^{t+\delta} ds' \hat{V}^{\boldsymbol{\lambda}\dagger}(s') \hat{\rho}_B]\hat{\sigma}_{mn}\hat{\rho}^{\boldsymbol{\lambda}}_S(t)\hat{\sigma}_{m'n'}^\dagger.
    \end{array}
    \label{eq:gme-aux}
\end{equation}
The r.h.s. term of the first line cancels out since the $\hat{\sigma}_{nn}\hat{\rho}^{\boldsymbol{\lambda}}_S(t)\hat{\sigma}_{n'n'}^\dagger=(\hat{\rho}^{\boldsymbol{\lambda}}_S(t))_{nn'}|E_n\rangle\langle E_{n'}|$ and $\{|E_n\rangle\}_n$ is a basis of the system. To compute the other terms, we write, as in the section 6 of this supplemental material, the interaction Hamiltonian in the general form $\hat{V}(t)=\hat{A}(t)\otimes\hat{B}^\dagger(t)+\hat{A}^\dagger(t)\otimes\hat{B}(t)$, where $\hat{A}(t)=\sum_{mn}g_{mn}\hat{\sigma}_{mn}e^{-i\omega_{mn}t}$ ($g_{mn}\in\mathbb{C}$) acts on the system Hilbert space and $\hat{B}(t)$ on the bath. 
The coupling Hamiltonian with counting fields reads (see also \eqref{eq:defdelta})
\begin{equation}
\begin{array}{ll}
 \hat{V}^{\boldsymbol{\lambda}}(t)  &=\sum_{mn}g_{mn}\hat{\sigma}_{mn}e^{-\lambda\omega_{mn}}\hat{B}^\dagger_{-\lambda}(t) e^{-i\omega_{mn}t}+g_{mn}^*\hat{\sigma}^\dagger_{mn}e^{\lambda\omega_{mn}}\hat{B}_{\lambda}(t) e^{i\omega_{mn}t} ,
\end{array}
    \label{eq:v-lambda}
\end{equation}
where $\hat{B}_\lambda(t)=e^{\lambda\hat{H}_B/2}\hat{B}(t)e^{-\lambda\hat{H}_B/2}$.
Using the above decomposition, and using the fact that $\{\hat{\sigma}_{mn}\}$ form an orthogonal basis, and $\sum_{mn}Tr_S[\hat{\sigma}_{mn}]\hat{\sigma}_{mn}=\mathds{I}$, we can re-write \eqref{eq:gme-aux} as
\begin{equation}
 \hspace{-1cm}   \begin{array}{l}
\mathcal{L}_{\boldsymbol{\lambda}}(\hat{\rho}_s(t))=\\
\\
\frac{1}{\delta_0}\sum\limits_{mn,m'n'}\int_t^{t+\delta_0} ds\int_t^{t+\delta_0} ds' Tr_B[ \hat{B}^\dagger_{-\lambda}(s)\hat{\rho}_B \hat{B}_{-\lambda}(s')]e^{-\lambda(\omega_{mn}+\omega_{m'n'})/2}e^{-i(\omega_{mn}s-\omega_{m'n'}s')}g_{mn}g^*_{m'n'}\hat{\sigma}_{mn}\hat{\rho}_S(t)\hat{\sigma}_{m'n'}^\dagger\\
    \\
    +\frac{1}{\delta_0}\sum\limits_{mn,m'n'}\int_t^{t+\delta_0} ds\int_t^{t+\delta_0} ds' Tr_B[ \hat{B}_{\lambda}(s)\hat{\rho}_B \hat{B}^\dagger_{\lambda}(s')]e^{\lambda(\omega_{mn}+\omega_{m'n'})/2}e^{i(\omega_{mn}s-\omega_{m'n'}s')}g^*_{mn}g_{m'n'}\hat{\sigma}^\dagger_{mn}\hat{\rho}_S(t)\hat{\sigma}_{m'n'}\\
    \\
    -\frac{1}{2}\frac{1}{\delta_0}\sum\limits_{mn,m'n'} \int_t^{t+\delta} ds \int_t^{s} ds' Tr_B[\hat{B}_{-\lambda}^\dagger(s)\hat{B}_\lambda(s')\hat{\rho}_B]e^{-\lambda(\omega_{mn}-\omega_{m'n'})/2}e^{-i(\omega_{mn}s-\omega_{m'n'}s')}g_{mn}g^*_{m'n'}\hat{\sigma}_{mn}\hat{\sigma}^\dagger_{m'n'}\hat{\rho}_S(t)\\
    \\
    -\frac{1}{2}\frac{1}{\delta_0}\sum\limits_{mn,m'n'} \int_t^{t+\delta} ds \int_t^{s} ds' Tr_B[\hat{B}_\lambda(s)\hat{B}_{-\lambda}^\dagger(s')\hat{\rho}_B]e^{\lambda(\omega_{mn}-\omega_{m'n'})/2}e^{i(\omega_{mn}s-\omega_{m'n'}s')}g_{mn}^*g_{m'n'}\hat{\sigma}^\dagger_{mn}\hat{\sigma}_{m'n'}\hat{\rho}_S(t)\\
    \\
    -\frac{1}{2}\frac{1}{\delta_0}\sum\limits_{mn,m'n'} \int_t^{t+\delta} ds \int_s^{t+\delta_0} ds' Tr_B[\hat{B}_\lambda^\dagger(s)\hat{B}_{-\lambda}(s')\hat{\rho}_B]e^{\lambda(\omega_{mn}-\omega_{m'n'})/2}e^{-i(\omega_{mn}s-\omega_{m'n'}s')}g_{mn}g^*_{m'n'}\hat{\rho}_S(t)\hat{\sigma}_{mn}\hat{\sigma}^\dagger_{m'n'}\\
    \\
    -\frac{1}{2}\frac{1}{\delta_0}\sum\limits_{mn,m'n'} \int_t^{t+\delta_0} ds \int_s^{t+\delta_0} ds' Tr_B[\hat{B}_{-\lambda}(s)\hat{B}_\lambda^\dagger(s')\hat{\rho}_B]e^{-\lambda(\omega_{mn}-\omega_{m'n'})/2}e^{i(\omega_{mn}s-\omega_{m'n'}s')}g^*_{mn}g_{m'n'}\hat{\rho}_S(t)\hat{\sigma}^\dagger_{mn}\hat{\sigma}_{m'n'}.
    \end{array}
    \label{eq:gme-aux-2}
\end{equation}
For the last two lines, we used the orthogonality property: $Tr_S[\hat{\sigma}^\dagger_{mn}\hat{\sigma}_{pq}\hat{\sigma}^\dagger_{p'q'}]\neq 0$ iff $\hat{\sigma}_{mn}=\hat{\sigma}_{pq}\hat{\sigma}^\dagger_{p'q'}$, and relabelled $\hat{\sigma}_{pq}\hat{\sigma}^\dagger_{p'q'}$ by  $\hat{\sigma}_{mn}\hat{\sigma}^\dagger_{m'n'}$. Notice that the tilted bath correlation functions in the last four lines are in fact $\lambda$ independent. This follows from the fact that $[\hat{\rho}_B,\hat{H}_B]=0$ and the definition $\hat{B}_\lambda(t)=e^{\lambda\hat{H}_B/2}\hat{B}(t)e^{-\lambda\hat{H}_B/2}$.

Let's now examine when the average energy conservation is satisfied, i.e., when do we have $\partial_\lambda Tr_S[\mathcal{L}_{\boldsymbol{\lambda}}(\hat{\rho}^{\boldsymbol{\lambda}}_S(t))]=0$. To do so, we compute directly the trace of the r.h.s. of \eqref{eq:gme-aux-2} and take the derivative in $\lambda$. We begin by noticing that the last four lines in \eqref{eq:gme-aux-2} cancel out. Indeed, since $\delta_0$ is much larger than the correlation time of the bath, the time integrals of the last four lines can be approximated by
\begin{equation}
    \begin{array}{l}
\frac{1}{\delta_0}\int\limits_t^{t+\delta_0} ds \int\limits_t^{s} ds' Tr_B[\hat{B}^\dagger(s)\hat{B}(s')\hat{\rho}_B]e^{-i(\omega_{mn}s-\omega_{m'n'}s')} = e^{-i(\omega_{mn}-\omega_{m'n'})(t+\delta_0/2)}\mbox{sc}\left(\frac{(\omega_{mn}-\omega_{m'n'})\delta_0}{2}\right)\Gamma_-(\omega_{m'n'}),\\
\\
\frac{1}{\delta_0}\int\limits_t^{t+\delta_0} ds \int\limits_s^{t+\delta_0} ds' Tr_B[\hat{B}^\dagger(s)\hat{B}(s')\hat{\rho}_B]e^{-i(\omega_{mn}s-\omega_{m'n'}s')} =  e^{-i(\omega_{mn}-\omega_{m'n'})(t+\delta_0/2)}\mbox{sc}\left(\frac{(\omega_{mn}-\omega_{m'n'})\delta_0}{2}\right)\Gamma_-(\omega_{m'n'}),
    \end{array}
    \label{eq:LS-coefs}
\end{equation}
where $\Gamma_-(\omega)=\int_0^{+\infty}Tr_B[\hat{B}^\dagger(t)\hat{B}(0)]e^{i\omega t} dt$ and sc is a short hand notation for $sinc$. The integrals with $Tr_B[\hat{B}(s)\hat{B}^\dagger(s')\hat{\rho}_B]$ have the same form with $\Gamma_-(\omega)$ being replaced by $\Gamma_+(\omega)=\int_0^{+\infty}Tr_B[\hat{B}(t)\hat{B}(0)^\dagger]e^{i\omega t} dt$. Hence, the coefficients of lines 3 and 5 in \eqref{eq:gme-aux-2} are identical, as well as the coefficients of lines 4 and 6. Since the sign of $\lambda$ alternates, the last four lines cancel out when we take the trace and the derivative in $\lambda$. We are left with the first two lines. It is convenient at this point to take the continuous limit for the baths, which is justified since we assume the heat baths to be at equilibrium. We therefore write $\hat{B}$ in the form
\begin{equation}
    \hat{B}(t)=\int_0^\Omega d\omega \sqrt{J(\omega)}\hat{b}(\omega)e^{-i\omega t},
\end{equation}
with $\hat{b},\hat{b}^\dagger$ bosonic annihilation and creation operators obeying the Bose Einstein statistics ($Tr_B[\hat{b}(\omega)^\dagger\hat{b}(\omega)\hat{\rho}_B]=1/(e^{\beta\omega}-1)=n_B(\omega)$,  $Tr_B[\hat{b}(\omega)\hat{b}^\dagger(\omega)\hat{\rho}_B]=n_B(\omega)+1$), $J$ the spectral density and $\Omega$ a cutoff frequency.  Writing explicitly
\begin{equation}
\begin{array}{ll}
    Tr_S[\hat{B}_{-\lambda}^\dagger(s)\hat{\rho}_B \hat{B}_{-\lambda}(s')]&=\int_0^\Omega d\omega J(\omega)(n_B(\omega)+1) e^{\lambda\omega}e^{i\omega( s-s')},\\
    \\
    Tr_S[\hat{B}_\lambda(s)\hat{\rho}_B \hat{B}^\dagger_{\lambda}(s')]&=\int_0^\Omega d\omega J(\omega)n_B(\omega) e^{-\lambda\omega}e^{-i\omega( s-s')},
\end{array}
\end{equation}
we obtain, after re-injecting the above expression in \eqref{eq:gme-aux-2} and taking the derivative in $\lambda$ of the trace, the condition
\begin{equation}
\hspace{-1cm}    \begin{array}{l}
\sum\limits_{mn,m'n'} Tr_S[\hat{\sigma}_{m'n'}^\dagger\hat{\sigma}_{mn}\hat{\rho}_S(t)]\int_0^\Omega \frac{2\omega -(\omega_{mn}+\omega_{m'n'})}{2} d\omega J(\omega)(n_B(\omega)+1) e^{-i(\omega_{mn}-\omega_{m'n'})\delta_0/2}\delta_0^2\mbox{sc}\left(\frac{\omega-\omega_{mn}}{2}\delta_0\right)\mbox{sc}\left(\frac{\omega-\omega_{m'n'}}{2}\delta_0\right)\\
\\
+\sum\limits_{mn,m'n'}Tr_S[\hat{\sigma}_{m'n'}\hat{\sigma}_{mn}^\dagger\hat{\rho}_S(t)]\int_0^\Omega \frac{2\omega -(\omega_{mn}+\omega_{m'n'})}{2} d\omega J(\omega)n_B(\omega) e^{i(\omega_{mn}-\omega_{m'n'})\delta_0/2}\delta_0^2\mbox{sc}\left(\frac{\omega-\omega_{mn}}{2}\delta_0\right)\mbox{sc}\left(\frac{\omega-\omega_{m'n'}}{2}\delta_0\right)=0.
    \end{array}
\end{equation}
Since the above equality must hold for every $\beta$, the two sums must individually vanish, and since it must hold for every $\hat{\rho}_S$, each coefficient must vanish. Hence,  the condition $\partial_\lambda Tr_S[\mathcal{L}_{\boldsymbol{\lambda}}(\hat{\rho}^{\boldsymbol{\lambda}}_S(t)]|_{\lambda=0}=0$ boils down to the condition (23),
\begin{equation}
    \begin{array}{ll}
\int d\omega\, R_\pm(\omega) \frac{2\omega-\omega_{nm}-\omega_{n'm'}}{2} \mbox{sc}\left(\frac{(\omega-\omega_{nm})\delta_0}{2}\right)\mbox{sc}\left(\frac{(\omega-\omega_{n'm'})\delta_0}{2}\right) =0,
    \end{array}
\end{equation}
where here $R_\pm(\omega)=J(\omega)(n_B(\omega)+\frac{1}{2}\pm\frac{1}{2})$. In the case of multiple, additive baths, the above condition needs to be satisfied for each bath.

\medskip

We now give an example of a QME satisfying the GQDB and the average energy conservation, but not the strict energy conservation (15). Starting from the general form of the tilted map in (9), with $\boldsymbol{\lambda}=(\lambda_S,\lambda_B)$ we consider the weak coupling case, 
%
%
and follow the same steps as above. When computing the time integrals, we make the following approximation:
\begin{equation}
\begin{array}{ll}
  \frac{1}{\delta_0}  \int d\omega R_\pm(\omega) \delta_0^2\mbox{sc}\left(\frac{(\omega-\omega_{nm})\delta_0}{2}\right)\mbox{sc}\left(\frac{(\omega-\omega_{n'm'})\delta_0}{2}\right) &\approx \int d\omega R_\pm(\omega) \delta_0\mbox{sc}\left(\frac{(2\omega-\omega_{nm}-\omega_{n'm'})\delta_0}{2}\right)\\
  \\
  &\approx \left\{
  \begin{array}{ll}
  R_\pm\left(\frac{\omega_{mn}+\omega_{m'n'}}{2}\right) &\mbox{if $|\omega_{mn}-\omega_{m'n'}|<\delta_0^{-1}$}\\
  0 & \mbox{if $|\omega_{mn}-\omega_{m'n'}|>\delta_0^{-1}$}
  \end{array}\right.
\end{array}
\end{equation}
and, for the terms in \eqref{eq:LS-coefs}, 
\begin{equation}
e^{\pm i(\omega_{mn}-\omega_{m'n'})\delta_0/2}\mbox{sc}\left(\frac{(\omega_{mn}-\omega_{m'n'})\delta_0}{2}\right)\Gamma_\pm(\omega_{m'n'})\approx \left\{
  \begin{array}{ll}
  \Gamma_\pm\left(\frac{\omega_{mn}+\omega_{m'n'}}{2}\right) &\mbox{if $|\omega_{mn}-\omega_{m'n'}|<\delta_0^{-1}$}\\
  0 & \mbox{if $|\omega_{mn}-\omega_{m'n'}|>\delta_0^{-1}$}
  \end{array}\right. \, .
\end{equation}
The imaginary parts of $\Gamma_\pm\left(\frac{\omega_{mn}+\omega_{m'n'}}{2}\right) $ are regrouped to form a Lamb shift term $\hat{H}_{LS}^{\boldsymbol{\lambda}}$ while the real parts constitute the anti commutator $\hat{G}_{\boldsymbol{\lambda}}$ of the dissipator. As explained earlier, the bath counting field $\lambda_B$ vanishes in these two terms, and only $\lambda_S$ remains. Going back to the Schrodinger picture, this finally leads to the following tilted superoperator,
\begin{equation}
\mathcal{L}_{\boldsymbol{\lambda}}(\hat{\rho}^{\boldsymbol{\lambda}}_S(t))=-i(\hat{H}_{LS}^{\lambda_S,0}\hat{\rho}^{\boldsymbol{\lambda}}_S(t)-\hat{\rho}^{\boldsymbol{\lambda}}_S(t)\hat{H}_{LS}^{-\lambda_S,0})+(\hat{G}_{\lambda_S,0}\hat{\rho}^{\boldsymbol{\lambda}}_S(t)-\hat{\rho}^{\boldsymbol{\lambda}}_S(t)\hat{G}_{-\lambda_S,0})+\mathcal{J}_{\boldsymbol{\lambda}}\hat{\rho}^{\boldsymbol{\lambda}}_S(t) \, ,
\label{eq:qme-gen}
\end{equation}
with
\begin{equation}
    \begin{array}{l}
\hat{H}_{LS}^{\lambda_S,0}=\frac{1}{2}\sum\limits_{mn,m'n'}I_{+,mn}^{m'n'}e^{-\lambda_S(\omega_{mn}-\omega_{m'n'})}\hat{A}^\dagger_{mn}\hat{A}_{m'n'}+I_{-,mn}^{m'n'}e^{\lambda_S(\omega_{mn}-\omega_{m'n'})}\hat{A}_{mn}\hat{A}^\dagger_{m'n'} \, ,\\
\\
\hat{G}_{\lambda_S,0}=-\frac{1}{2}\sum\limits_{mn,m'n'}R_{+,mn}^{m'n'}e^{-\lambda_S(\omega_{mn}-\omega_{m'n'})}\hat{A}^\dagger_{mn}\hat{A}_{m'n'}+R_{-,mn}^{m'n'}e^{\lambda_S(\omega_{mn}-\omega_{m'n'})}\hat{A}_{mn}\hat{A}^\dagger_{m'n'} \, ,\\
\\
\mathcal{J}_{\boldsymbol{\lambda}}\hat{\rho}^{\boldsymbol{\lambda}}_S(t)= \sum\limits_{mn,m'n'}R_{+,mn}^{m'n'}e^{-\frac{(\lambda_S-\lambda_B)(\omega_{mn}+\omega_{m'n'})}{2}}\hat{A}_{mn}\hat{\rho}^{\boldsymbol{\lambda}}_S(t)\hat{A}^\dagger_{m'n'}+R_{-,mn}^{m'n'}e^{\frac{(\lambda_S-\lambda_B)(\omega_{mn}+\omega_{m'n'})}{2}}\hat{A}^\dagger_{mn}\hat{\rho}^{\boldsymbol{\lambda}}_S(t)\hat{A}_{m'n'} \, ,
    \end{array}
\end{equation}
where $\hat{A}_{mn}=g_{mn}\hat{\sigma}_{mn}$, and where $R_{\pm,mn}^{m'n'}$ is a short hand notation for $R_\pm\left(\frac{\omega_{mn}+\omega_{m'n'}}{2}\right)$. Since the matrix of coefficients $[R_{\pm,mn}^{m'n'}]$ is positive, we can symmetrize the superoperator and write it in a Lindblad form. One can readily check that \eqref{eq:qme-gen} satisfies the GQDB (13) and the average energy conservation (23), hence (18), but not the strict energy conservation (15).

\section{8 - Restoring the GQDB using approximations of the Redfield ME}\label{app:red-sym}

Let us go back to the tilted Redfield ME \eqref{eq:red-lambda}. To obtain a tilted QME, of Lindblad form when the counting fields are set to zero, we may extend the procedures developed in \cite{wacker2018,ptaszynski2019thermodynamics,farina2019,mccauley2020accurate,nathan2020universal, potts2021} -- which consist in performing approximations to the Redfield ME in order to restore its positivity -- to the tilted Redfield ME \eqref{eq:red-lambda}.  When the counting fields are set to zero, these different procedures use the idea that, if the functions $\Gamma^\alpha_\pm$ are smooth enough, we may replace the terms $(\Gamma^\alpha_\pm(\omega_{m'n'},0)^*+\Gamma^\alpha_\pm(\omega_{mn},0))$ by a function symmetric in $\omega_{mn},\omega_{m'n'}$.
This "symmetrization" procedure can be also done at the level of the real and imaginary parts of $\Gamma^{\alpha}_{\pm}$, that is  $R_\pm^\alpha(\omega_{mn},0)$ and  $I_\pm^\alpha(\omega_{mn},0)$. This allows to transform \eqref{eq:red-lambda} (when $\boldsymbol{\lambda_B=0}$) in a QME that preserves the positivity of the dynamics. Extending these symmetrization procedures to the tilted generator \eqref{eq:red-lambda}, one would replace the terms $(\Gamma^\alpha_\pm(\omega_{m'n'},\lambda_\alpha)^*+\Gamma^\alpha_\pm(\omega_{mn},\lambda_\alpha))$ by symmetric functions in $\omega_{mn},\omega_{m'n'}$. Provided that the symmetry \eqref{eq:sym-gamma} is maintained, one would obtain a tilted QME, positive when $\lambda_S=0$ and $\boldsymbol{\lambda_B=0}$, and satisfying the GQDB.

To give an example, let's consider the following procedure, inspired by \cite{mccauley2020accurate}. Without loss of generality, we consider the case of a single bath. We drop the index $\alpha$, and note $\lambda_B$ the counting field on the bath. Assuming that the Lamb shift terms $I_\pm(\omega_{mn},0)$ are larger than the damping rates $R_\pm(\omega_{mn},0)$ (the reasoning is similar in the opposite case), let us define the parameter $\epsilon=\mbox{min}_{mn,m'n'}\{\sqrt{I_\pm(\omega_{mn},0)I_\pm(\omega_{m'n'},0)}\}$ where the minimum is taken over the frequencies associated to the couple of jump operators in $\hat{H}_{LS}$. We then perform the substitution, for any $\omega_{mn},\omega_{m'n'}$ in \eqref{eq:LS-red},
\begin{equation}
    \begin{array}{ll}
R_\pm(\omega_{mn},\lambda_B)&\approx 
\left\{ 
\begin{array}{ll}
\sqrt{R_\pm(\omega_{mn},\lambda_B)R_\pm(\omega_{m'n'},\lambda_B)} & \mbox{if $|\omega_{mn}-\omega_{m'n'}|< \epsilon$}\\
\\
0 & \mbox{otherwise,}
\end{array}
\right. \\
\\
I_\pm(\omega_{mn},\lambda_B)&\approx
\left\{ 
\begin{array}{ll}
\mbox{sign}(I_\pm(\omega_{mn},\lambda_B)) \sqrt{|I_\pm(\omega_{mn},\lambda_B)I_\pm^\alpha(\omega_{m'n'},\lambda_B)|} & \mbox{if $|\omega_{mn}-\omega_{m'n'}|< \epsilon$}\\
\\
0 & \mbox{otherwise.}
\end{array}
\right. 
\end{array}
    \label{eq:sqrt}
\end{equation}
The substitution \eqref{eq:sqrt} is valid provided that $\Gamma_\pm^\alpha$ varies smoothly over intervals of width $\epsilon$, and induces a correction of order $|R^{\alpha'}_\pm/R^\alpha_\pm|\epsilon$, $|I^{\alpha'}_\pm/I^\alpha_\pm|\epsilon$.
Finally, in order to identify $\hat{H}_{LS}$ with a Lamb shift term which commutes with $\hat{H}_S$, we perform the secular approximation on $\hat{H}_{LS}$. Adding the counting field $\lambda_S$ on the system, we obtain
\begin{equation}
\begin{array}{l}
\mathcal{L}^{sym}_{\lambda_S,\lambda_B}(\hat{\rho}_S(t))=-i[\hat{H}_S+\hat{H}_{LS},\hat{\rho}_S(t)]  \\
\\
+ \sum\limits_{mn,m'n'} \sqrt{R_+(\omega_{mn})R_+(\omega_{m'n'})}e^{-(\lambda_S-\lambda_B)(\omega_{mn}+\omega_{m'n'})/2}\hat{A}_{mn}\hat{\rho}_S(t)\hat{A}_{m'n'}^\dagger\\
\\
-\frac{1}{2}\sum\limits_{mn,m'n'} \sqrt{R_+(\omega_{mn})R_+(\omega_{m'n'})}(e^{\lambda_S(\omega_{mn}-\omega_{m'n'})/2}\hat{A}^\dagger_{mn}\hat{A}_{m'n'}\hat{\rho}_S(t)+e^{-\lambda_S(\omega_{mn}-\omega_{m'n'})/2}\hat{\rho}_S(t)\hat{A}^\dagger_{mn}\hat{A}_{m'n'})\\
\\
+ \sum\limits_{mn,m'n'} \sqrt{R_-(\omega_{mn})R_-(\omega_{m'n'})}e^{(\lambda_S-\lambda_B)(\omega_{mn}+\omega_{m'n'})/2}\hat{A}^\dagger_{mn}\hat{\rho}_S(t)\hat{A}_{m'n'}\\
\\
-\frac{1}{2}\sum\limits_{mn,m'n'} \sqrt{R_-(\omega_{mn})R_-(\omega_{m'n'})}(e^{-\lambda_S(\omega_{mn}-\omega_{m'n'})/2}\hat{A}_{mn}\hat{A}^\dagger_{m'n'}\hat{\rho}_S(t)+e^{-\lambda_S(\omega_{mn}-\omega_{m'n'})/2}\hat{\rho}_S(t)\hat{A}_{mn}\hat{A}^\dagger_{m'n'}),
\end{array}
\label{eq:tilted-sym}
\end{equation}
where, to alleviate the notation, we wrote $R(\omega_{mn})=R(\omega_{mn},0)$.
As announced, the resulting $\mathcal{L}^{sym}_{0,\lambda_B}$ satisfies the GQDB. In addition, the average energy conservation is also satisfied, i.e., $\partial_\lambda Tr_S[\mathcal{L}^{sym}_{\lambda,\lambda}(\hat{\rho}_S)]=0$ for all $\hat{\rho}_S$. However, since $\mathcal{L}^{sym}_{\lambda,\lambda}$ is still $\lambda$ dependent, the strict energy conservation (15) is not satisfied. We will use this procedure to compute the heat flow between a three level system and a heat bath and compare it with an exact numerical simulation at the end of this supplemental material.


\section{9 - Numerical study of a three level system}\label{app:3level}

\begin{figure}[h]
    \centering
    \includegraphics[scale=0.4]{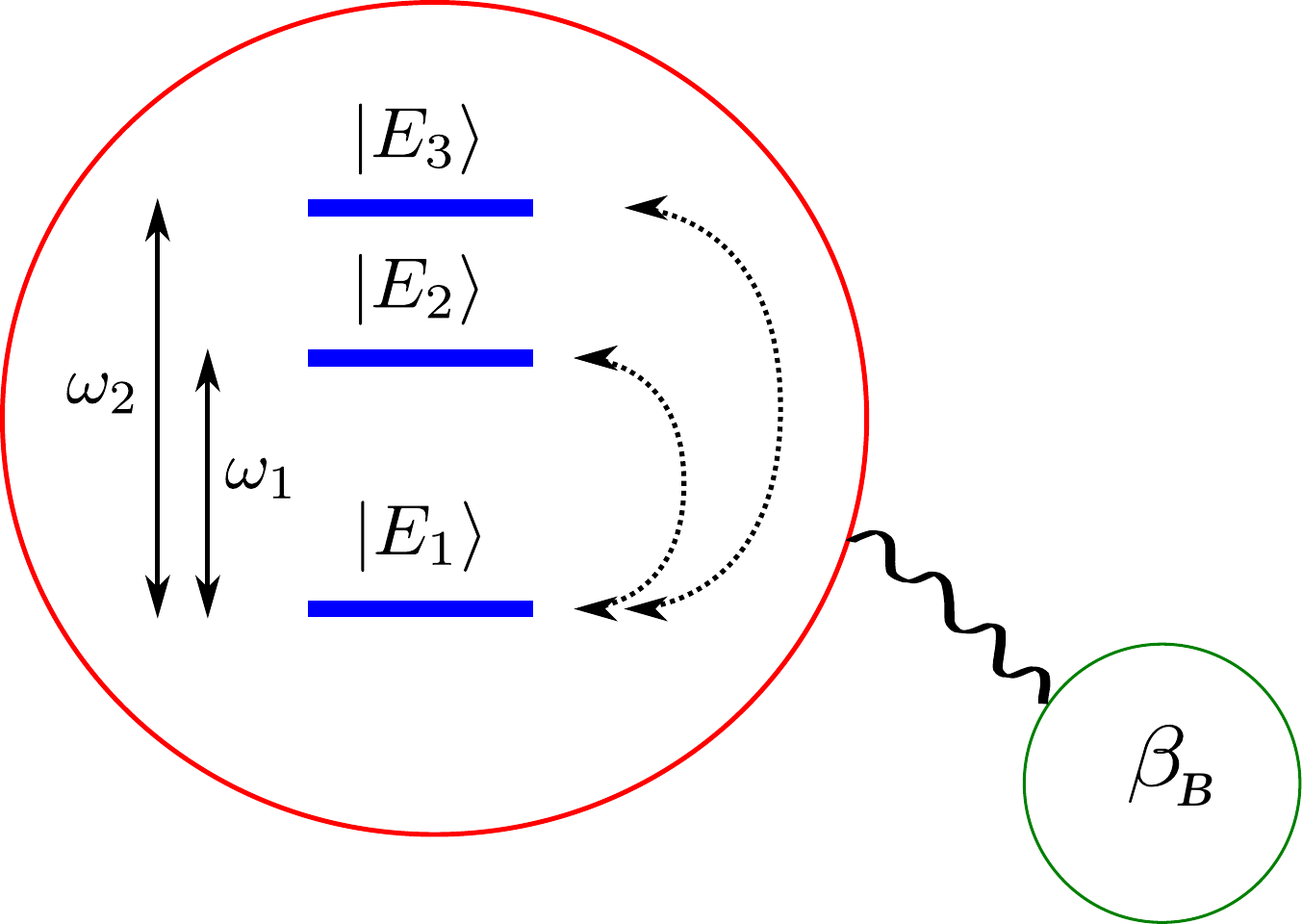}
    \caption{Schematic representation of the setup analysed numerically. A three level system is weakly coupled to a thermal bath at temperature $\beta_B^{-1}$. The only jumps allowed are between the ground state $|E_1\rangle$ and the excited states $|E_1\rangle$, $|E_2\rangle$. }
    \label{fig:3level}
\end{figure}

\begin{figure}
    \centering
    \includegraphics[scale=0.6]{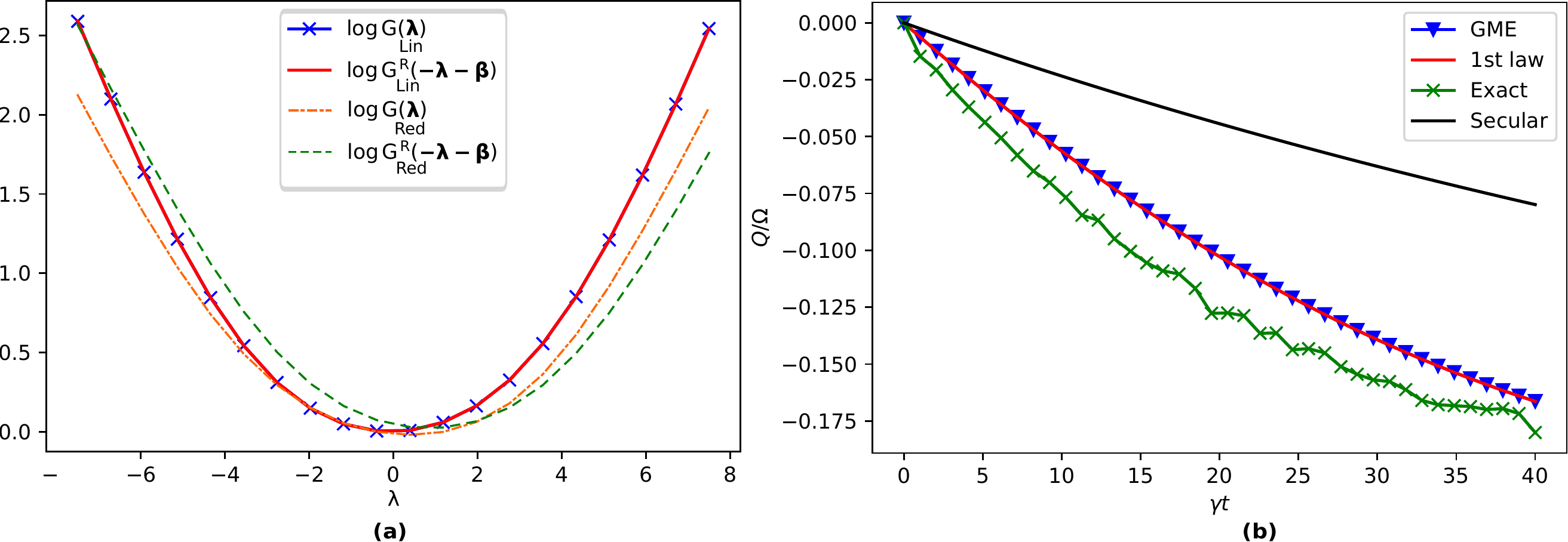}
    \caption{{\bf (a)} Log of the MGF for the QME of Linblad form ($G_{Lin}$) and for the Redfield ME ($G_{Red}$) describing the a three-level system (see Fig.\ref{fig:3level}) coupled to a thermal bath, with $\boldsymbol{\lambda}=(\lambda, -\lambda)$. The MGFs are evaluated at fixed time $t$ in the transient regime (since $t$ is fixed, we drop it in the legend to alleviate the notations). The fluctuation theorem is satisfied for the QME of Lindblad form, but not for the Redfield ME. {\bf (b)} Heat $Q$ leaked from the bath to the system in the setup Fig.\ref{fig:3level}. The red curve represents the system energy variations, which, if the first law is satisfied, should correspond to the heat, $ Q=Tr_S[\hat{H}_S(\hat{\rho}_S(t)-\hat{\rho}_S(0))]$, while the blue curve shows the heat computed from the MGF, $\partial_{\lambda_B}G(t,-\lambda_B)|_{\lambda_B=0}$. The black curve gives the heat computed with the secular approximation, and the green curve is the heat computed from an exact simulation.}
    \label{fig:cgf}
\end{figure}

Consider the three level system sketched in Fig.\ref{fig:3level} where ($\hbar =1$): $\hat{H}_B =\int\limits_0^\Omega d\omega\, \omega\, \hat{b}^\dagger(\omega) \hat{b}(\omega)$,
$\hat{V}=\gamma (\hat{A}\otimes\hat{B}^\dagger+\hat{A}^\dagger\otimes\hat{B})$,
$\hat{A} =\sum\limits_{j=1}^2g_j\hat{\sigma}_j$, $\hat{B}=\int\limits_0^\Omega d\omega \sqrt{J(\omega)}\hat{b}(\omega)$. Here $\hat{b}^\dagger(\omega), \hat{b}(\omega)$ are bosonic creation and annihiliation operators, $\Omega$ a frequency cutoff, $\gamma$ a dimensionless coupling constant, and $\hat{\sigma}_{j}$ jump operators between the excited eigenstates of the system and the ground state: $\hat{\sigma}_1=|0\rangle\langle 1|$, $\hat{\sigma}_2=|0\rangle\langle 2|$.
Following the method presented in the section 8 of this supplemental material, we derive a tilted superoperator of the form \eqref{eq:tilted-sym}, which we solve numerically in order to compute the heat exchanges with the bath.
The parameters are set to $\gamma=0.2$, $\beta_B=5/\Omega$, $g1=g2=\sqrt{\Omega}$, $E_2=0.1\Omega-1/2\delta_0$, $E_3=0.1\Omega+1/2\delta_0$, where $\delta_0$ is chosen as the geometric average of the bath and system relaxation times. The system is initially in the state $\hat{\rho}_S(0)=(|E_2\rangle+\sqrt{3}|E_3\rangle)(\langle E_2|+\sqrt{3}\langle E_3|)/4$.
The steady state for the model, $\hat{\rho}_{ss}$, is not a Gibbs state ($\hat{\rho}_{G}$), and has small surviving coherences between the two excited states; we find that the mismatch between the steady state populations and the Gibbs distribution is of the same order as the steady state coherences, $\langle j|\hat{\rho}_{ss}-\rho_G|j\rangle \sim\langle 1| \hat{\rho}_{ss} |2\rangle \sim 10^{-4}$. Despite that, FT (5) is satisfied for this model while it breaks if we adopt a description based on the Redfield ME, see Fig.\ref{fig:cgf}.a. 

We also check that energy conservation is satisfied on average, namely that $Q=Tr_S[\hat{H}_S(\hat{\rho}_S(t)-\hat{\rho}_S(0)]$, see Fig.\ref{fig:cgf}.b. Finally, we compare the accuracy of the symmetrized Lindblad ME \eqref{eq:tilted-sym} with that of the secular Lindblad equation, by computing the heat flow predicted by these two equations and comparing it to the exact heat flow, see Fig.\ref{fig:cgf}.b. In order to compute the exact heat flow numerically, we model the bath Hamiltonian $\hat{H}_B$ by a diagonal matrix with $N=1500$ equally
spaced eigenvalues distributed between $−0.5$ and $0.5$. The interaction Hamiltonian $\hat{V}$ is chosen of the form
$\hat{V}=\gamma(\hat{A}+\hat{A}^\dagger)\otimes \hat{R} $, where $\hat{R}=\hat{X}/4\sqrt{ N}$ with $\hat{X}$ a Gaussian orthogonal random matrix of size $N$ with probability density
proportional to $e^{− Tr(\hat{X}^2)/4 }$ \cite{mehta2004randmat}. The exact heat flow is then obtained from the heat MGF for the exact dynamics. As seen from the Fig.\ref{fig:cgf}.b, in the present case where the energy differences of the system are nearly degenerate, the symmetrized Lindblad QME is more accurate than the secular Lindblad QME.

\section{10 - Consistency in the weak coupling limit}\label{app:approx-2ndlaw}

As opposed to secular QMEs, the QMEs of Lindblad form obtained from perturbative developments of (9) or from approximations of the Redfield ME do not necessarily admit the Gibbs state as a steady state when the system is coupled to a single bath. This results from the fact that these QMEs couple the dynamics of the diagonal terms of the density matrix (populations) and the dynamics of the off diagonal terms (coherences), unlike the secular QMEs. The fact that the Gibbs state is not a steady state seemingly leads to a breakdown of the second law (19). Indeed, if we initiate the system in a Gibbs state $\hat{\rho}_S(0)=\hat{\rho}_G$, and using (18), we obtain $\langle\Sigma\rangle=-D(\hat{\rho}_S(t)||\hat{\rho}_G)$, which is negative if $\hat{\rho}_S(t)\neq\hat{\rho}_G$. Let us now show that $D(\hat{\rho}_S(t)||\hat{\rho}_G)$ is then negligible w.r.t. the approximations made to derive the symmetrized QME from the exact dynamics, i.e., that the second law holds.

To see this, we consider a system coupled to a single bath at temperature $\beta^{-1}$. We assume that the total density matrix is factorized at time $t$, $\hat{\rho}(t)=\hat{\rho}_S(t)\otimes\hat{\rho}_B(t)$ with $\hat{\rho}_B(t)=e^{-\beta\hat{H}_B}/Z_B$. The dynamics of the total system is governed by a unitary operator, leading, after a time $\delta_0$, to the density matrix $\hat{\rho}(t+\delta_0)$. Following the steps of \cite{esposito2010entropy}, we then decompose the relative entropy between the total density matrix and  $\hat{\rho}_S(t+\delta_0)\otimes\hat{\rho}_B$, where $\hat{\rho}_S(t+\delta_0)=Tr_B[\hat{\rho}(t+\delta_0)]$ as
\begin{equation}
\begin{array}{ll}
  D(\hat{\rho}(t+\delta_0)||\hat{\rho}_S(t+\delta_0)\otimes\hat{\rho}_B) &=  -S_{tot}(t+\delta_0)+S(t+\delta_0)-Tr_B[\hat{\rho}_B(t+\delta_0)\log\hat{\rho}_B(t)]\\
  \\
  &= -S(t)+Tr_B[\hat{\rho}_B(t)\log\hat{\rho}_B(t)]+S(t+\delta_0)-Tr_B[\hat{\rho}_B(t+\delta_0)\log\hat{\rho}_B(t)]\\
  \\
  &=S(t+\delta_0) -S(t) +\beta Tr[\hat{H}_B(\rho(t+\delta_0)-\rho(t)] ,
\end{array}
  \label{eq:sigma-tot}
\end{equation}
where $S_{tot}$ is the total entropy, which is preserved, $S_{tot}(t+\delta_0)=S_{tot}(t)$, and where as in the main text $S(t)=-Tr_S[\hat{\rho}_S(t)\log\hat{\rho}_S(t)]$. Since a relative entropy is always positive, we have
\begin{equation}
S(t+\delta_0) -S(t) +\beta Tr[\hat{H}_B(\rho(t+\delta_0)-\rho(t)] \geq 0.
\label{eq:entropy-aux}
\end{equation}
Note that $Tr[\hat{H}_B(\hat{\rho}(t+\delta_0)-\hat{\rho}(t))]=-Q$ where $Q$ is the heat leaked from the bath. When an average energy conservation is satisfied, the bath energy variations translate exactly into variations of the system energy,
\begin{equation}
    Tr[\hat{H}_B(\hat{\rho}(t+\delta_0)-\hat{\rho}(t))] = -Tr_S[\hat{H}_S(\hat{\rho}_S(t+\delta_0)-\hat{\rho}_S(t))] \, ,
\end{equation}
and, after replacing in \eqref{eq:entropy-aux}, we obtain
\begin{equation}
    S(t+\delta_0) -S(t) - \beta Q\geq 0,
    \label{eq:entropy-aux2}
\end{equation}
where now $Q=Tr_S[\hat{H}_S(\hat{\rho}_S(t+\delta_0)-\hat{\rho}_S(t))] $. To make a connection with (19) from the main text, notice that the system density matrix is here formally defined as $\hat{\rho}_S(t)=Tr_B[\hat{\rho}(t)]$. Since the QME is obtained from the unitary dynamics by tracing out the bath degrees of freedom and making additional approximations, we can write $Tr_B[\hat{\rho}(t)] = \hat{\rho}_S^{QME}+\hat{\epsilon}$, where $\hat{\rho}_S^{QME}$ denotes the density matrix obtained by solving the QME, and $\hat{\epsilon}$ accounts for the approximations. Replacing in \eqref{eq:entropy-aux2}, and using the subscript $QME$ to distinguish the quantities derived using $\hat{\rho}^{QME}_S$ rather that the exact solution $\hat{\rho}_S(t)=Tr_B[\hat{\rho}(t)]$, we obtain
\begin{equation}
    S^{QME}(t+\delta_0) -S^{QME}(t) - \beta Q^{QME} + \mathcal{O}(\hat{\epsilon}) \geq 0.
\end{equation}
By virtue of the semigroup hypothesis, we can repeat the same reasoning for any time interval larger than $\delta_0$, and obtain
\begin{equation}
    \Delta S^{QME} - \beta Q^{QME} + \mathcal{O}(\hat{\epsilon}) \geq 0.
    \label{eq:2nd-qme}
\end{equation}
The quantity $\Delta S^{QME} - \beta Q^{QME} $, which is the entropy production computed using the QME, is therefore positive up to fluctuations of order $\hat{\epsilon}$.

To make the argument more concrete, let's show that $D(\hat{\rho}_{ss}||\hat{\rho}_G)$ is in fact of the order $\hat{V}^3$, hence negligible, in the context of QMEs derived weak coupling limit using a perturbative development to second order in $\hat{V}$. Without loss of generality, we assume that the system Hamiltonian $\hat{H}_S$ is time independent. Recall that we pointed out, at the beginning of the letter, that in order to write the first law as $W=\Delta E_S-Q$ with $\Delta E_S=Tr_S[\hat{H}_S(\hat{\rho}_S(t)-\hat{\rho}_S(0))]$ and $Q$ the heat leaving the bath, we need to require that the coupling is switched on after the initial measurement and switched off before the final one. Otherwise, we need to include the contribution of the coupling to the system energy variation. When we require the average energy conservation for the QME, we should also require the coupling to be switched on/off at the beginning and end of each time interval $[t,t+\delta_0]$, starting with $[0,\delta_0]$; instead, we are assuming the coupling to stay constant between the initial and final switched on/off and are neglecting its contribution. Therefore, the exact system energy variations are given by $\Delta E_S=Tr[(\hat{H}_S+\hat{V})(\hat{\rho}(t)-\hat{\rho}(0))]$. Since we assumed $\hat{H}_S$ and $\hat{V}$ to be time independent, $W=0$ and hence the heat leaked from the bath is $Q=Tr[(\hat{H}_S+\hat{V})(\hat{\rho}(t)-\hat{\rho}(0))]$. Re-injecting in \eqref{eq:sigma-tot}, we obtain
\begin{equation}
    \Delta S -\beta Tr[(\hat{H}_S+\hat{V})(\hat{\rho}(t)-\hat{\rho}(0))] = \langle\Sigma^{QME}\rangle -\beta Tr[\hat{V}(\hat{\rho}(t)-\hat{\rho}(0))]\geq 0,
\end{equation}
where 
\begin{equation}
 \langle\Sigma^{QME}\rangle=\Delta S^{QME} -\beta Tr_S[\hat{H}_S(\hat{\rho}^{QME}_S(t)-\hat{\rho}^{QME}_S(0))]   
\end{equation}
is the entropy production predicted by the QME. Let's now show that $Tr[\hat{V}(\hat{\rho}(t)-\hat{\rho}(0))]$ is negligible in the weak coupling limit. Consider $Tr[\hat{V}(\hat{\rho}(t+\delta_0)-\hat{\rho}(t))]$. The semigroup hypothesis allows to write
\begin{equation}
    Tr[\hat{V}(\hat{\rho}(t+\delta_0)-\hat{\rho}(t))] = Tr[\hat{V}(\hat{M}(t+\delta_0,t)-\mathds{I})\hat{\rho}_S(t)],
\end{equation}
where $\hat{M}=\hat{M}_{\boldsymbol{\lambda}}$ as defined in (8) with $\boldsymbol{\lambda}=\boldsymbol{0}$. As showed in the section 7 of this supplemental material, in the weak coupling limit, $(\hat{M}(t+\delta_0,t)-\mathds{I})$ consists of terms of order $\mathcal{O}(\hat{V}^2)$, hence $Tr[\hat{V}(\hat{\rho}(t+\delta_0)-\hat{\rho}(t))]=\mathcal{O}(\hat{V}^3) $. We can apply the same reasoning for the Redfield ME, by considering $d_t Tr[\hat{V}(\hat{\rho}(t)-\hat{\rho}(0))]=Tr[\hat{V}(\dot{\hat{\rho}}(t)]$ and, following the textbook derivation of the Redfield equation \cite{breuer2002theory}, replace 
\begin{equation}
    \dot{\hat{\rho}}(t)=-\int_0^t ds [\hat{V}(t),[\hat{V}(s),\hat{\rho}(s)]]\approx -\int_0^t ds [\hat{V}(t),[\hat{V}(s),\hat{\rho}_S(s)\otimes\hat{\rho}_B]],
\end{equation}
where we use the weak coupling approximation to write $\hat{\rho}(t)\approx\hat{\rho}_S(t)\otimes\hat{\rho}_B$. Hence, once again $Tr[\hat{V}(\hat{\rho}(t+\delta_0)-\hat{\rho}(t))]=\mathcal{O}(\hat{V}^3) $.

Finally, let us point out that, if a QME satisfies an average energy conservation, then the first law is expressed as
\begin{equation}
    \Delta E_S^{QME}=Q^{QME}.
    \label{eq:1st-qme}
\end{equation}
As a sanity check, let us prove the identity \eqref{eq:1st-qme} explicitly for the symmetrized Lindblad QME derived in the section 6 of this supplemental material, without resorting to counting fields. More precisely, we prove that 
\begin{equation}
   -d_t Tr[\hat{H}_B(\hat{\rho}(t)] = d_tTr[\hat{H}_S(\hat{\rho}(t)],
\end{equation}
when $d_t T_B[\hat{\rho}(t)]$ is computed using the QME. Without loss of generality, we assume that the system is coupled to a single bath. Let us begin with the l.h.s. term. As in the section 6 of this supplemental material, we begin with deriving the Redfield ME, but keeping $\hat{H}_B$:
\begin{equation}
\begin{array}{ll}
-d_t Tr[\hat{H}_B(\hat{\rho}(t)]&=Tr_S[\int_0^{+\infty}ds Tr_B[\hat{H}_B[\hat{A}(t)\otimes\hat{B}^\dagger(t),[\hat{A}^\dagger(t-s)\otimes\hat{B}(t-s),\hat{\rho}_S(t)\otimes\hat{\rho}_B]] ] + h.c.\\
\\
&=\int_0^{+\infty}ds Tr_S[\hat{A}(t)\hat{A}^\dagger(t-s)\hat{\rho}_S(t)] Tr_B[\hat{H}_B\hat{B}^\dagger(t)\hat{B}(t-s)\hat{\rho}_B-\hat{H}_B\hat{B}(t-s)\hat{\rho}_B \hat{B}^\dagger(t)]\\
\\
&+\int_0^{+\infty}ds Tr_S[\hat{A}^\dagger(t-s)\hat{A}(t)\hat{\rho}_S(t)]Tr_B[\hat{H}_B\hat{\rho}_B\hat{B}(t-s)\hat{B}^\dagger(t)-\hat{H}_B\hat{B}^\dagger(t)\hat{\rho}_B \hat{B}(t-s)] + h.c.\\
\\
&=\int_0^{+\infty}ds Tr_S[\hat{A}(t)\hat{A}^\dagger(t-s)\hat{\rho}_S(t)] Tr_B[\hat{B}^\dagger(t)[\hat{B}(t-s),\hat{H}_B]\hat{\rho}_B]\\
\\
&+\int_0^{+\infty}ds Tr_S[\hat{A}^\dagger(t-s)\hat{A}(t)\hat{\rho}_S(t)]Tr_B[\hat{B}(t-s)[\hat{B}^\dagger(t),\hat{H}_B]\hat{\rho}_B] + h.c.
\end{array}
\label{eq:check-1}
\end{equation}
On the other hand,
\begin{equation}
    \begin{array}{ll}
d_t Tr[\hat{H}_S(\hat{\rho}(t)]=&Tr_S[\int_0^{+\infty}ds Tr_B[\hat{H}_S[\hat{A}(t)\otimes\hat{B}^\dagger(t),[\hat{A}^\dagger(t-s)\otimes\hat{B}(t-s),\hat{\rho}_S(t)\otimes\hat{\rho}_B]] ] + h.c.\\
\\
&=-\int_0^{+\infty}ds Tr_S[\hat{A}(t)[\hat{A}^\dagger(t-s),\hat{H}_S]\hat{\rho}_S]Tr_B[\hat{B}^\dagger(t) \hat{B}(t-s)\hat{\rho}_B]\\
\\
&-\int_0^{+\infty}ds Tr_S[\hat{A}^\dagger(t-s)[\hat{A}(t),\hat{H}_S]\hat{\rho}_S]Tr_B[\hat{B}(t-s) \hat{B}^\dagger(t)\hat{\rho}_B].
    \end{array}
    \label{eq:check-2}
\end{equation}
Assuming, as in the section 6 of this supplemental material, that the bath is a collection of harmonic oscillators linearly coupled with the system: $\hat{B}_\alpha=\sum_k j_{\alpha,k} \hat{b}_\alpha(\omega_{\alpha,k})$, s.t. $[\hat{H}_B,\hat{b}_\alpha(\omega_{\alpha,k})]=-\omega_{\alpha,k}\hat{b}_\alpha(\omega_k)$ and $[\hat{H}_\alpha,\hat{b}_\alpha^\dagger(\omega_{\alpha,k})]=\omega_{\alpha,k}\hat{b}_\alpha^\dagger(\omega_{\alpha,k})$, we have
\begin{equation}
    \begin{array}{ll}
\int_0^{+\infty}ds e^{-i\omega_{mn} s} Tr_B[\hat{B}^\dagger(t)[\hat{B}(t-s),\hat{H}_B]\hat{\rho}_B] &= \omega_{mn}\int_0^{+\infty}ds e^{-i\omega_{mn} s} Tr_B[\hat{B}^\dagger(t)\hat{B}(t-s)\hat{\rho}_B], \\
\\
\int_0^{+\infty}ds e^{-i\omega_{mn} s} Tr_B[\hat{B}(t-s)[\hat{B}^\dagger(t),\hat{H}_B]\hat{\rho}_B] &= -\omega_{mn}\int_0^{+\infty}ds e^{-i\omega_{mn} s} Tr_B[\hat{B}(t-s)\hat{B}^\dagger(t)\hat{\rho}_B] \, .
    \end{array}
\end{equation}
Replacing in \eqref{eq:check-1}, and recalling that $\hat{A}=\sum_{mn}g_{mn}\hat{\sigma}_{mn}$ with $[\hat{\sigma}_{mn},\hat{H}_S]=\omega_{mn}\sigma_{mn}$, we find that the terms on the l.h.s. of \eqref{eq:check-1} and \eqref{eq:check-2} are equal. Applying the symmetrization of the coefficients, we obtain \eqref{eq:1st-qme}, which we wanted to prove.

\end{document}